\documentclass[aps,showpacs,showkeys,longbibliography,nofootinbib]{revtex4-1}
\usepackage[latin1]{inputenc} 
\usepackage{amssymb}
\usepackage{amsfonts} 
\usepackage{amsthm} 
\usepackage{amsmath,latexsym} 
\usepackage[makeroom]{cancel}
\usepackage{epsfig}
\usepackage{hyperref}
\usepackage{xcolor}
\usepackage{bbold}
\usepackage[T1]{fontenc}
\usepackage{young}
\usepackage{youngtab}
\usepackage[boxsize=1.07em]{ytableau}
\usepackage{mathtools}
\usepackage{braket}
\usepackage{slashed}
\usepackage{physics}

\makeatletter
\let\cat@comma@active\@empty
\makeatother

\def\ic{\mathrm{i}}

\def \bc {\begin{center}}
\def \ec {\end{center}}
\def \bi {\begin{itemize}}
\def \ei {\end{itemize}}

\def \ba {\begin{array}}
\def \ea {\end{array}}

\def \bea {\begin{eqnarray}}
\def \eea {\end{eqnarray}}

\def \be {\begin{equation}}
\def \ee {\end{equation}}

\newcommand{\la}{\langle}
\newcommand{\ra}{\rangle}

\def \um {\frac{1}{2}}

\def\tr {\mathrm{tr}}

\def\flux {L}

\def\mb{\mathbb{m}}
\def\adhw{A^\dag_{\mathrm{hw}}}
\def\mbhw{\mathbb{m}_{\mathrm{hw}}}
\def\mblw{\mathbb{m}_{\mathrm{lw}}}

\def\rmu{\mathrm{U}}
\def\rmsu{\mathrm{SU}}

\newcommand\quelle[1]{{%
		\unskip\nobreak\hfil\penalty50
		\hskip2em\hbox{}\nobreak\hfil\textbf{#1}%
		\parfillskip=0pt \finalhyphendemerits=0 \par}}

\newcommand\mydots{\hbox to 0.9em{.\hss.\hss.}}

\newtheorem{thm}{Theorem}[section]

\newtheorem{lem}[thm]{Lemma}
\newtheorem{prop}[thm]{Proposition}

\theoremstyle{remark}

\begin{document}

Published in Symmetry  14(5), 872 (2022): \href{https://doi.org/10.3390/sym14050872}{https://doi.org/10.3390/sym14050872}

\title{Hilbert space structure of the low energy sector of U(N) quantum Hall ferromagnets and their classical limit}

\author{Manuel Calixto}
\email{Corresponding author: calixto@ugr.es}
\affiliation{Department of Applied Mathematics, University of  Granada,
Fuentenueva s/n, 18071 Granada, Spain}
\affiliation{Institute Carlos I of Theoretical and Computational Physics (iC1), University of  Granada,
Fuentenueva s/n, 18071 Granada, Spain}
\author{Alberto Mayorgas}
\email{albmayrey97@ugr.es}
\affiliation{Department of Applied Mathematics, University of  Granada,
Fuentenueva s/n, 18071 Granada, Spain}
\author{Julio Guerrero}
\email{jguerrer@ujaen.es}
\affiliation{Department of Mathematics, University of Jaen, Campus Las Lagunillas s/n, 23071 Jaen, Spain}
\affiliation{Institute Carlos I of Theoretical and Computational Physics (iC1), University of  Granada,
Fuentenueva s/n, 18071 Granada, Spain}

\date{\today}

\begin{abstract}
Using the Lieb-Mattis ordering theorem of electronic energy levels, we identify the Hilbert space of the low energy sector of $\rmu(N)$ quantum Hall/Heisenberg ferromagnets at filling factor $M$ for $L$ Landau/lattice sites 
with the carrier space of irreducible representations of $\rmu(N)$ described by rectangular Young tableaux of $M$ rows 
and $L$ columns, and associated to Grassmannian phase spaces $\rmu(N)/\rmu(M)\times \rmu(N-M)$. 
We embed this $N$-component fermion mixture in Fock space through a Schwinger-Jordan (boson and fermion) representation of $\rmu(N)$-spin operators. 
We provide different realizations of basis vectors using Young diagrams, Gelfand-Tsetlin patterns and Fock states (for an electron/flux occupation number in the fermionic/bosonic representation). 
$\rmu(N)$-spin operator matrix elements in the Gelfand-Tsetlin basis are explicitly given. Coherent state excitations above the ground state are computed and labeled by complex $(N-M)\times M$ matrix points $Z$ on the Grassmannian 
phase space. They adopt the form of a $\rmu(N)$ displaced/rotated highest-weight vector, or a multinomial Bose-Einstein condensate in the flux occupation number representation.  
Replacing $\rmu(N)$-spin operators by their expectation values in a Grassmannian coherent state allows for a semi-classical treatment of the low energy (long wavelength) $\rmu(N)$-spin-wave coherent excitations (skyrmions) of 
$\rmu(N)$ quantum Hall ferromagnets in terms of Grasmannian nonlinear sigma models.
\end{abstract}

%
%
%
%
%

\maketitle

\section{Introduction}\label{intro}

The breakthrough in the development of new quantum technologies requires mathematical modeling and an adequate theoretical framework for the study of the underlying nuclear, atomic, molecular, optical and condensed matter systems.  
Algebraic, numerical, analytical and topological mathematical tools for dealing with complex many-body quantum systems are necessary to analyze their properties. In particular, for the understanding of new and exotic topological quantum phases 
of matter [characterized by topological numbers like Chern, Pontryagin, Skyrmion \eqref{Pontryagineq}, etc. and other winding numbers] and their exploitation for technological applications. Indeed, the discovery of new quantum phases of matter 
(mainly of a topological nature), their classification, analysis and understanding, is a very hot/topical subject. 
High-temperature superconductors and an emergent category two-dimensional materials provide new types of topological phases, sometimes characterized by exotic electronic (edge) states and currents remarkably robust to impurities and thermal fluctuations. 
Quantum Hall effect provides the paradigmatic example of a topological phase, but dispersion-less edge currents also appear in the absence of magnetic field, for example, in some graphene analogues (silicene, germanene, etc) with a strong 
spin-orbit coupling. Two-dimensional topological insulators (see \cite{TI2} for a text book, \cite{RevModPhys.82.3045,RevModPhys.83.1057} for reviews and \cite{Kou2017} for progress and prospects) were predicted theoretically by 
Kane and Mele \cite{KaneMele05} using a two-dimensional graphene-like material model with spin-orbit interaction. 
They were first proposed  \cite{BernevigHgTe} and  observed experimentally \cite{KonigScience2007} in mercury cadmium telluride (HgTe/CdTe) 
semiconductor quantum wells and later in other materials. Another rapidly developing field has to do with topological quantum computation; see \cite{KITAEV2003} for Kitaev's original proposal, \cite{pachos_2012} for a text book 
and \cite{Sarma2015} for a current perspective on Majorana zero modes. 
Topological quantum computation is an approach to fault-tolerant quantum computation in which the unitary quantum gates result from the braiding of certain topological quantum objects called ``anyons''. 
Topological degrees of freedom promise to encode decoherence-resistant and scalable quantum information. For example, magnetic skyrmions are promising for technological applications, including spintronics and neuromorphic computing. 
They  might be used as information carriers in future advanced memories, logic gates and computing devices (see \cite{Zhang2015,Zhang2015b} and \cite{Zhang2016} for bilayer systems). 
The creation and transmission of an isolated magnetic skyrmion in thin films is a key for future skyrmionics, which utilizes skyrmions as information carriers in advanced memories, 
logic gates and computing devices \cite{Zhang2015,Zhang2015b}. Therefore, a deeper fundamental/theoretical study of models related to this subject is justified by its future use in quantum technologies.

In this article we concentrate on the study of systems of interacting $N$-component fermions. Traditionally, the paradigmatic  case for electrons is $N=2$ (spin 1/2), extensible to  $N=3$ (flavor, color) components for leptons and quarks 
(see \cite{Jacak2022} for high energy consequences of topological quantum effects),  $N=4$ (spin-isospin) 
components in nuclear physics, etc. The  subject of $\rmsu(N)$ fermions has been recently further fueled in condensed matter physics 
by the fact that $\rmsu(N)$ symmetries can be extended to larger $N$ in ultracold atomic  gases (see e.g. the text books \cite{pethick_smith_2008,Lewenstein_2012} and \cite{Cazalilla_2014} for a review).  
For example,  fermionic alkaline-earth atomic gases trapped in optical lattices realize the $\rmsu(N)$ generalization of the Hubbard model \cite{PhysRevLett.92.170403,Cazalilla_2009}. 
Exciting recent advances in cooling, trapping and manipulating more and more complex systems of this kind, make Feynman's original ideas about the simulation of 
quantum systems and quantum information processing increasingly possible.

Here we want to revisit and deepen the particular subject of $\rmu(N)$ quantum Hall (Heisenberg-like) ferromagnets (QHF). 
As it is briefly reviewed in Appendix \ref{exchangeapp}, the exchange interaction for $N$-component planar electrons in a perpendicular magnetic field adopts the form of a $\rmu(N)$ QHF Hamiltonian 
\be
H=-\mathcal{J}\sum_{\langle \alpha,\beta\rangle}\sum_{i,j=1}^N S_{ij}(\alpha)S_{ji}(\beta), 
\ee
on a square lattice when written in terms of $\rmu(N)$-spin operators $S_{ij}(\alpha)=c^\dag_i(\alpha) c_j(\alpha)$ realized in terms of 
creation $c^\dag_i(\alpha)$ and annihilation $c_i(\alpha)$ operators of an electron 
with component $i,j=1,\dots,N$ in a given Landau/lattice site $\alpha$ of a given Landau level (namely, the lowest one). The sum over $\langle \alpha,\beta\rangle$ extends over 
all near-neighbor Landau/lattice sites, and $\mathcal{J}$ is the exchange coupling constant (the spin stiffness for the XY model). 
Electrons become multicomponent when, for example, in addition to the usual spin components $\uparrow$ and $\downarrow$, they acquire extra ``pseudospin''  internal 
components associated: (a) with layer (for a multilayer arrangement), (b) with valley (like in graphene and other 2D Dirac materials), (c) with sub-lattice, etc.  In addition, 
multilayer arrays introduce extra components (``flavors'') to the electron  and much richness, so that the unitary group U$(N)$ also plays a fundamental role here. 
For example, twisted bilayer (and trilayer) graphene for ``magic'' angles exhibit interesting superconducting properties \cite{Bistritzer12233,Cao2018}. In the case of  a bilayer quantum Hall system in the lowest Landau level, 
one Landau site can accommodate $N=4$ internal   states $|i\ra, i=1,2,3,4$ (let us call them fermion  ``flavors/components'', in general); more schematically  
\be |1\rangle=|\uparrow t\rangle,\quad |2\ra=|\uparrow b\ra,\quad  |3\ra=|\downarrow t\ra,\quad |4\ra=|\downarrow b\ra,\label{iso4}\ee
where $t$ and $b$ make reference to the ``top'' and ``bottom'' layers, respectively. Since the electron field has $N=4$ degenerate components, the bilayer system possesses an underlying $\rmu(4)$ symmetry. 
Likewise, the $\ell$-layer case carries a $\rmu(2\ell)$ symmetry (see next Section \ref{ferrosec} for more details).

For $N$-component electrons, the Pauli exclusion principle allows $M\leq N$ electrons per Landau/lattice site (the filling factor). 
Selecting a ground state $\Phi_0$ ($|0\ra_\mathrm{F}$ denotes the Fock vacuum) 
\be
|\Phi_0\ra=\Pi_{\alpha=1}^{\flux}\Pi_{i=1}^M c_i^\dag(\alpha)|0\ra_\mathrm{F},\label{GS}
\ee
which fills all $\flux$ Landau sites with the first $M$ internal levels $i=1,\dots,M\leq N$ [i.e., for integer  filling factor $M$], spontaneously breaks the $\rmu(N)$ symmetry 
since a general unitary transformation mixes the first $M$ ``spontaneously chosen'' 
occupied internal levels with the $N-M$ unoccupied ones. The ground state $|\Phi_0\ra$ is  still invariant under the stability subgroup $\rmu(M)\times \rmu(N-M)$ of transformations among the $M$ occupied levels and 
the $N-M$ unoccupied levels, respectively.  Therefore, the transformations that do not leave  $|\Phi_0\ra$ invariant 
are parametrized by the Grassmannian coset $\mathbb{G}^N_M=\rmu(N)/\rmu(M)\times \rmu(N-M)$, which reduces to the well known 
sphere $\mathbb{S}^2=\rmu(2)/\rmu(1)\times \rmu(1)$ for $N=2$ spin  components and $M=1$ electron per Landau site. 
The kind of irreducible representations (IRs) of $\rmu(N)$ related to Grassmann phase spaces $\mathbb{G}^N_M$ are those described by rectangular Young tableaux of $M$ rows and $\flux$ columns, 
where $\flux$ labels the corresponding IR, just as spin $s$ does for $\rmsu(2)$. 

The objective of this article is to describe the carrier Hilbert space   associated 
with these $\rmu(N)$ representations, their coherent states (see e.g. the standard text books \cite{Perelomov,Gazeaubook}), and the classical limit. 
In the classical limit $\flux\to\infty$ (large $\rmu(N)$-spin representations), the collective operators $S_{ij}$ become c-numbers 
(their coherent state expectation values, to be more precise), 
and the low energy (long wavelength) $\rmu(N)$-spin-wave coherent excitations are named ``skyrmions'' (see e.g. some recent books and thesis \cite{Seki-Mochizuki_Skyrmion_2016,JungHoonHan_Skyrmion_2017,Finocchio-Panagopoulos_Skyrmion_2021,Zang_Thesis_Skyrmion_2018}).
These coherent excitations  turn out to be described by 
a ferromagnetic order parameter associated to this spontaneous symmetry breaking and  labeled by 
$(N-M)\times M$ complex matrices $Z$ parametrizing the  complex Grassmannian 
manifold $\mathbb{G}_M^N$ (see later on Section \ref{Grassec} for more information about its structure). Actually, the classical dynamics associated to these $\rmsu(N)$ quantum spin chains  
can be described by a Grassmannian nonlinear sigma model (NL$\sigma$M)  \cite{AffleckNPB257,AffleckNPB265,AffleckNPB305,Sachdev,Sachdev2,Arovas},  generalizing the $\rmsu(2)$ NL$\sigma$M for 
the continuum dynamics of  Heisenberg (anti)ferromagnets \cite{HaldanePLA93,HaldanePLA93-2,HaldanePLA93-3}. In references like \cite{AffleckNPB305,Sachdev}, $N$ represents the number of fermion ``flavors'', 
whereas $\flux$ is referred to as the number of ``colours'' $n_c$.

The organization of the paper is the following. In Section  \ref{ferrosec} we motivate the description of low energy sectors of  $\rmu(N)$ QHF by representations linked rectangular Young tableaux, using the 
Lieb-Mattis ordering of electronic energy levels based on the pouring principle for Young tableaux. 
In Section \ref{Qsec} we develop this idea and construct the Hilbert space of $M\flux$ $N$-component fermions occupying $\flux$ Landau sites (integer filling factor $M$) 
making use of a bosonic realization of  the $\rmu(N)$-spin collective operators $S_{ij}$ acting on  Fock space states. 
The whole construction relies on the definition of a highest-weight (ground) state in Sec. \ref{hwsec}, a ``boson condensate'' version of the  ground state \eqref{GS}. 
We provide a representation of basis vectors in terms of Young tableaux, 
Gelfand-Tsetlin  vectors and Fock (boson and fermion) states \ref{recyoungel}. The monolayer $N=2$ case at filling factor $M=1$, the bilayer $N=4$ case at filling factor $M=2$ and 
the trilayer $N=6$ case at $M=3$ are worked out as particular examples. General Hilbert-space dimension formulas are provided in Sec. \ref{gendimsec}. 
Matrix elements of the $\rmu(N)$ physical  operators are provided in Sec. \ref{matsec}, together with the spectrum of Casimir operators, 
paying special attention to the quadratic Casimir operator since it is related with the exchange interaction Hamiltonian at low energies. Section \ref{Grassec} is devoted to the discussion of Grassmannian coherent states 
and the expectation values of $\rmu(N)$-spin collective operators. $\rmu(N)$ coherent states  can be seen as coherent excitations above the 
highest weight (ground) state in the form of Bose-Einstein condensates. Coherent states are essential to 
discuss the classical limit of large $\flux$ representations of $\rmu(N)$ QHF in terms of NL$\sigma$Ms on Grassmannian manifolds $\mathbb{G}^N_M$, of which we also comment in the second half 
of Section \ref{Grassec}. The last Section \ref{conclusec} is devoted to 
conclusions and outlook. For completeness, and to be as self-contained as possible, we include in Appendix \ref{exchangeapp}  a  brief remind on the derivation of $\rmu(N)$ QHF 
models from first principle (two-body exchange) interactions. The proof of propositions  \ref{rectdomin} and \ref{groundstate} is given in Appendices  \ref{rectdominproof} and \ref{hwapp}, 
respectively. A more detailed relation between Gelfand-Tsetlin and Fock states is left for the Appendix \ref{appGelfandFock}. The spin-pseudospin structure of basis states for 
bilayer $\rmu(4)$ QHF at filling factor $M=2$ is made explicit in the Appendix \ref{spinpspinapp}. Explicit particular expressions of $\rmu(N)$-spin matrices for $N=4, M=1, L=1, 2$ and $N=4, 
M=2, L=1$ are given in Appendix \ref{appmatrixelements}. Finally, general considerations about the highest weight state for  Young tableaux of arbitrary shape are given in Appendix \ref{appsec}.

\section{U(N) ferromagnetism and Lieb-Mattis ordering of electronic energy levels}\label{ferrosec}

Let us denote by $\mathcal H_{N}^\alpha[1^M]$ the $\tbinom{N}{M}$-dimensional carrier Hilbert space at site $\alpha$ of the fully antisymmetric IR  of $\rmu(N)$ described by the Young frames/diagrams of shape $[1^M]$, 
that is, with $M$ boxes on a single column. This is a convenient way of graphically representing $\rmu(N)$ (and symmetric group $\mathfrak{S}_P$) invariant subspaces, i.e.,  by Young diagrams  of 
$P$ boxes/particles
\begin{equation}\label{youngdiagram}
	\overbrace{
		\begin{gathered}
			\begin{ytableau}
				~ &...&...&...&...&...&~\\
				: &:& : & : &:\\
				~&...&~
			\end{ytableau}
		\end{gathered}
	}^{h_1}
\end{equation}
of shape  $h=[h_1,\dots,h_N]$, with  $h_1\geq \dots \geq h_N$, $h_i$ the number of boxes in row $i=1,\dots,N$ and $h_1+\dots +h_N=P$ the total number of particles. This is why $h$ is also called a partition of $P$. 
We sometimes use the shorthand 
$[h,\stackrel{M}{\dots},h,0,\dots,0]=[h^M]$, obviating zero-box rows.  Basis vectors  of $\mathcal H_{N}^\alpha[1^M]$ are the  $M$-particle  Slater determinants (in Fock and  Young tableau notation)
\be
\Yvcentermath1  \Pi_{\mu=1}^M c_{i_\mu}^\dag(\alpha)|0\ra_\mathrm{F}=\young({{i_1}},:,{{i_M}}) 
\ee
obtained by filling out columns of the corresponding Young diagram with components $i_\mu\in\{1,\dots,N\}$ in strictly increasing order $i_1<\dots <i_M$. The ground state (``highest weight'') vector \eqref{GS} is just 
one example. One can see that there are exactly $\tbinom{N}{M}$ different arrangements of this kind 
(the dimension of $\mathcal H_{N}^\alpha[1^M]$). 

The Hilbert space of a $\rmu(N)$ QHF with $\flux$ Landau/lattice sites at integer filling factor $M$  is the $\tbinom{N}{M}^\flux$-dimensional $\flux$-fold 
tensor product space $\mathcal H_{N}^{\otimes\flux}[1^M]=\bigotimes_{\alpha=1}^\flux\mathcal H_{N}^\alpha[1^M]$.  In Young tableau notation
\be
\Yvcentermath1  \\[3pt] M\Bigg\{\young(\quad,:,\quad)  \:\:\otimes\:\: \stackrel{\flux\, \mathrm{times}}{\dots}  \:\:\otimes\:\:\young(\quad,:,\quad) \quad \leftrightarrow \quad  
[1^M]^{\otimes\flux}=[1^M]\otimes\stackrel{\flux}{\dots}\otimes [1^M]\,.\label{tensorprod}
\ee
This tensor product representation of $\rmu(N)$ is reducible and it decomposes into a direct sum of irreducible representations 
of different shapes. For example, the Clebsch-Gordan decomposition of a tensor product of $\flux=2$ IRs of $\rmu(N)$ of shape $[1^M]$, with filling factor $M=2\leq N\geq 4$, is represented by the following Young diagrams
\be
\Yvcentermath1   \young(\quad,\quad) \:\:\otimes\:\: \young(\quad,\quad)\:\:=\:\:\young(\quad\quad,\quad\quad)\:\:\oplus\:\: \young(\quad\quad,\quad,\quad)\:\:\oplus\:\:\young(\quad,\quad,\quad,\quad)  \quad 
\leftrightarrow \quad  [1^2]\otimes[1^2]=[2^2]\oplus [2,1^2]\oplus [1^4].\label{lambda2nu2}
\ee
The $P(=M\flux)$-particle ground state \eqref{GS} is a vector of  $\mathcal H_{N}^{\otimes\flux}[1^M]$. In particular, for filling factor $M=2$ and $\flux=2$ lattice sites, $|\Phi_0\ra$ is represented by the rectangular Young tableau
\be
\Yvcentermath1  \Pi_{\alpha=1}^2\Pi_{i=1}^2 c_i^\dag(\alpha)|0\ra_\mathrm{F}=\young(11,22), 
\ee
where rows are filled  in a non-decreasing order. One can see that, for $N=4$ electron components, there are 20 different Young tableau arrangements of this kind, which is the dimension of the IR of $\rmu(4)$ given by the rectangular Young diagram of 
shape $[2^2]$ (see later on Section \ref{gendimsec} for general dimension formulas). In fact, the corresponding dimensions for the tensor product decomposition \eqref{lambda2nu2} for $N=4$ is $6\times 6=20+15+1$. Note that the 
ground state \eqref{GS} is invariant under permutation of lattice sites $\alpha$ (look at the equations (\ref{cAnticommutators},~\ref{cCommutators})); therefore, it will always belong to  IRs of $\rmu(N)$ of rectangular shape 
\begin{equation}
	[\flux^M]= M
	\Bigg\{
	\overbrace{
		\begin{gathered}
			\begin{ytableau}
				~ &...&~\\
				:& : & : \\
				~&...&~
			\end{ytableau}
		\end{gathered}
	}^{\flux}
	\label{CGdecomp}
\end{equation}
arising in the Clebsch-Gordan decomposition of the tensor product \eqref{tensorprod}. The rectangular Young tableaux of shape $[\flux^M]$ are antisymmetric under the interchange of rows (electron components or ``flavors'')  
and symmetric under the interchange of columns (lattice sites or ``colors''). 

Let us show how Lieb-Mattis' theorem \cite{Lieb-Mattis_PR1962}, and some generalizations \cite{Decamp_PRR2020}, also confer ``dominance'' to the rectangular Young diagrams 
$[\flux^M]$ [like $[2^2]$ in \eqref{lambda2nu2}] over the rest of diagrams arising in the Clebsch-Gordan decomposition of \eqref{tensorprod}. 
The set of Young diagrams is partially ordered (not all $P$-particle diagrams can be compared for $P>5$)  by the so called ``dominance order'' $\succeq$, such that 
\be\label{DominanceIneq}
[h_1,\dots,h_N]\succeq [h_1',\dots,h_N']\leftrightarrow h_1+\dots+h_k\geq h_1'+\dots+h_k' \quad \forall k\in[1,N]\,.
\ee
It is said that $h$ dominates $h'$ or that $h'$ precedes $h$ ($h'\preceq h$). Intuitively, it means that one can go from $h$ to $h'$ by moving a certain number of boxes from
upper rows to lower rows, so that $h$ is ``more symmetric''. Lieb-Mattis' theorem  \cite{Lieb-Mattis_PR1962} talks about the ``pouring principle'', saying that $h'$ can be 
``poured into'' $h$.  The theorem states that, under general conditions on the symmetric Hamiltonian of the system, if $h'\preceq h$ then $E(h)<E(h')$ [$E(h)\leq E(h')$ for ``pathologic''  
potentials], with $E(h)$ the ground state energy inside each IR $h$ of $\rmu(N)$. From this, the following proposition can be demonstrated

\begin{prop}\label{rectdomin} All Young diagrams arising in the Clebsch-Gordan direct sum decomposition of the $\flux$-fold 
tensor product \eqref{tensorprod} can be pored into the rectangular Young tableaux of shape $[\flux^M]$. That is, the ground state for a  $\rmu(N)$ QHF at filling factor $M$ 
belongs to the carrier Hilbert space $\mathcal H_N[\flux^M]$ of the rectangular IR $[\flux^M]$ of  $\rmu(N)$  inside 
the total Hilbert space $\mathcal H_{N}^{\otimes\flux}[1^M]$.
\end{prop}
\noindent The proof is made in the Appendix \ref{rectdominproof}.
Note that states in $[\flux^M]$ are invariant under the permutation of lattice sites $\alpha=1,\dots,\flux$, thus becoming indistinguishable (``bosonized''). Another way of interpreting it is the following. Given the Fourier 
transform $\mathcal{S}_{ij}(q)=\sum_{\alpha=1}^\flux e^{\ic q\alpha} S_{ij}(\alpha)$ of $\rmu(N)$-spin operators, the long-wavelength (low momentum/energy $q\simeq 0$) ground state excitations are described by the collective operators 
$\mathcal{S}_{ij}(0)=\sum_{\alpha=1}^\flux S_{ij}(\alpha)$, which are invariant under site permutations $\alpha\leftrightarrow\alpha'$. Moreover, the low-energy long-wavelength semi-classical ($\flux\to\infty$) dynamics 
of $\rmu(N)$ QHF is described by a NL$\sigma$M which target space is the Grasmannian (the phase space associated to $\rmu(N)$ IRs with rectangular Young diagrams).

Once we have motivated/highlighted the dominant role of rectangular Young diagrams of shape $[\flux^M]$ at low energies, let us  explicitly construct these representations in a boson realization of $\rmu(N)$ generators, 
together with their associated coherent states labeled by matrix points $Z$ on the Grassmann phase space $\rmu(N)/\rmu(M)\times \rmu(N-M)$. 
These kind of representations have been studied in (mainly mathematically oriented) text books like \cite{Barut}, but rarely associated with the low energy sector of spin systems like the ones pursued in this article. 
This is why we think this discussion deserves attention.

\section{Low energy sector of U(N) quantum Hall ferromagnets  at filling factor M}\label{Qsec}

\subsection{Boson realization of U(N)-spin operators, Fock space, highest-weight state and ladder operators} \label{hwsec}

In the quantum Hall approach, each electron  occupies  on average  a  surface  area of $2\pi\ell_B^2$ (a Landau site, with $\ell_B$ the magnetic length) that  is  pierced  by  one magnetic flux  quantum $\phi_0=2\pi\hbar/e$ 
(see Appendix \ref{exchangeapp} for more information about this picture). 
This image allows a dual bosonic Schwinger realization  of $\rmu(N)$-spin operators 
\be
S_{ij}=\sum_{\mu=1}^Ma^\dag_{i\mu} a_{j\mu},\; i,j=1,\dots,N, \label{Aij}
\ee
this time in terms of creation $a^\dag_{i\mu}$ and annihilation $a_{j\mu}$ \emph{boson} operators of magnetic flux quanta attached to the electron 
 $\mu=1,\dots,M$ with  component  $i=1,\dots,N$ [we use Greek indices $\mu,\nu$ for electron labels to avoid confusion]. From the usual bosonic commutation relations $[a_{i\mu},a^\dag_{j\nu}]=\delta_{ij}\delta_{\mu\nu}$ we derive 
 the $\rmu(N)$-spin commutation relations
 \be
\left[S_{{ij}},S_{{kl}}\right]=\delta _{{jk}} S_{{il}} -\delta _{{il}} S_{{kj}},\label{commurel}
\ee
where $\delta_{jk}$ is the usual Kronecker delta. This bosonic picture is quite common in algebraic approaches to nuclear and molecular structure \cite{Casten1993,Frankvanisacker,Iachellolevine}, for example in the interacting boson model (IBM) 
\cite{iachello_arima_1987}. Therefore, we have a representation of $\rmu(N)$ in Fock space made of Fock states 
\be
|n\rangle=\frac{\prod_{i=1}^N\prod_{\mu=1}^M (a^\dag_{i\mu})^{n_{\mu i}}}{(\prod_{i=1}^N\prod_{\mu=1}^M n_{\mu i}!)^{1/2}}|0\rangle_\mathrm{F}.\label{Fockstate}
\ee
The exponent $n_{\mu i}$ of $a^\dag_{i\mu}$ indicates the number of Landau/lattice sites (flux quanta) available to the electron $\mu$ of flavor $i$ (that is, the occupancy number $a^\dag_{i\mu}a_{i\mu}$). We write  
$n_{\mu i}$ and not $n_{i\mu}$ because $\mu$ will later make reference to a row index of a Young diagram. Since $\rmu(N)$ IRs are finite-dimensional, we know that the representation of $\rmu(N)$ in Fock space must be reducible. In particular, 
$\rmu(N)$-spin operators conserve the total number of particles $C_1=\sum_{i=1}^N S_{ii}\sim M\flux$ [the linear Casimir operator of $\rmu(N)$]. According to Schur's lemma, for a $\rmu(N)$ IR, every operator acting on the representation space  and commuting with all $S_{ij}$ must be trivial 
(a multiple of the identity). Note that the operators 
\be
\Lambda_{\mu\nu}=\sum_{i=1}^N a^\dag_{i\mu}a_{i\nu},\quad \mu,\nu=1,\dots,M,
\ee
close a  $\rmu_\Lambda(M)$ Lie algebra, where we are writing the subscript $\Lambda$ to emphasize that this is different from all the other appearances of $\rmu(M)$ that are related to the ``first'' $S_{ij}$ generators ($i,j\leq M$) 
of $\rmu(M)$ as a subgroup of $\rmu(N)$. In fact,  the operators $\Lambda_{\mu\nu}$ are in general independent of 
$S_{ij}$, that is,  $\Lambda_{\mu\nu}$ can not be written in terms of $S_{ij}$ (except for $M=1$, when $\Lambda_{11}=C_1$) since they realize an independent Lie algebra. 
The operators $\Lambda_{\mu\nu}$ preserve the IR space and they commute with all the $\rmu(N)$-spin operators, i.e.  
\[ [S_{ij},\Lambda_{\mu\nu}]=\sum_{\mu'=1}^M\sum_{i'=1}^N[a^\dag_{i\mu'} a_{j\mu'},a^\dag_{i'\mu}a_{i'\nu}]=0. \]
Therefore, the operators $\Lambda_{\mu\nu}$ can be consistently imposed as constraints on Fock state vectors to reduce the representation of  $S_{ij}$ in a consistent manner. 
Actually, since each electron has $\flux$ Landau/lattice sites at its disposal (i.e., a total number of $\flux$ flux quanta),  then the constraint $\Lambda_{\mu\mu}|\psi_\mathrm{phys.}\ra=  \flux|\psi_\mathrm{phys.}\ra$ 
has to be imposed on physical states $|\psi_\mathrm{phys.}\ra$. This constraint is extended to $\Lambda_{\mu\nu}|\psi_\mathrm{phys.}\ra=  \flux\delta_{\mu\nu}|\psi_\mathrm{phys.}\ra$  for non-diagonal $\mu\not=\nu$ operators  
[see below in eq. \eqref{p1} for the case of the highest weight vector]. For the basis Fock states \eqref{Fockstate}, the constraint $\Lambda_{\mu\mu}|n\rangle=\flux |n\rangle$ in particular means that 
$\sum_{i=1}^N n_{\mu i}=\flux$, the total number of Landau/lattice sites available to electron $\mu$.

Our aim is to construct a state basis  of the 
Hilbert space $\mathcal H_N[L^M]\subset \mathcal H_{N,M}^{\otimes L}$,  carrying  the IR $[L^M]$ of $\rmu(N)$, given in terms of linear combinations 
of Fock states \eqref{Fockstate}. The Hilbert space $\mathcal H_N[L^M]$ can be constructed from the so called  ``highest weight'' HW (resp. lowest-weight)  vector $|\mbhw \rangle$ 
by applying lowering  $S_{ij}, i>j$ (resp. raising $S_{ij}, i<j$) operators (see below for a more detailed explanation). This procedure reminds the standard construction of $\rmsu(2)$ spin-$j$ (Dicke) 
states $\{|j,m\ra, m=-j,\dots,j\}$ from the highest (resp. lowest)  
weight state $|j,j\ra$ (resp.  $|j,-j\ra$) by applying ladder angular momentum operators $J_-$ (resp. $J_+$). Given a common eigenvector $|\psi_w\rangle$ of $S_{ii}, i=1,\dots,N$, its 
weight $w=[w_1,\dots,w_N]$ is made of the corresponding eigenvalues $w_i,  i=1,\dots,N$, which count the number of electrons with flavor/component $i$; therefore, $w_1+\dots +w_N=P=M\flux$, the total number of particles, which is the 
value of the linear Casimir operator $C_1=S_{11}+\dots+S_{NN}$ of $\rmu(N)$. Any other state  $|\psi_{w'}\rangle$ has lower weight $w'$ than $|\psi_w\rangle$ if  
the first non-vanishing coefficient of $w-w'$ is positive. It is clear that the highest weight must be  $W=[\flux,\stackrel{M}{\dots},\flux,0,\stackrel{N-M}{\dots},0]$, which can also be read from the shape  of the Young diagram $[L^M]$ 
(remember that we are discarding zeros). Let us state this in a more formal way. Before, 
for the sake of compact notation, we shall denote  by 
\be
A=\begin{pmatrix} a_{11}  &\dots & a_{1M}  \\  \vdots  & 
& \vdots  \\ a_{N1} &  \dots & a_{NM}\end{pmatrix}, \; A^\dag=\begin{pmatrix} a^\dag_{11}  &\dots & a^\dag_{N1}  \\  \vdots  & 
& \vdots  \\ a^\dag_{1M} &  \dots & a^\dag_{NM}\end{pmatrix},\label{aadag}
\ee
the $N\times M$ and $M\times N$ annihilation and creation operator matrices, respectively.

\begin{prop}\label{groundstate} Let $A^\dag_{\mathrm{hw}}$ be the $M\times M$ submatrix 
\be A^\dag_{\mathrm{hw}}=\begin{pmatrix} a^\dag_{11}  &\dots & a^\dag_{M1}  \\  \vdots  & 
& \vdots  \\ a^\dag_{1M} &  \dots & a^\dag_{MM}\end{pmatrix}
\ee
of $A^\dag$ in \eqref{aadag}, given by its first $M$ columns (the leading principal submatrix of order $M$). 
Then the  state 
\be
|\mbhw \rangle=\frac{\det(\adhw)^{\flux}}{\left(\prod_{p=1}^{M} (p)_\flux\right)^{1/2}}|0\ra_\mathrm{F},\quad (p)_\flux=p(p+1)\dots (p+\flux-1)\label{highestweight}\ee
satisfies the highest weight (HW) conditions:
\bea
\Lambda_{\mu\nu}|\mbhw \rangle&=&\flux\delta_{\mu\nu} |\mbhw \rangle, \; \mu,\nu=1,\dots,M,\label{p1}\\
S_{ij}|\mbhw \rangle&=&\left\{\begin{array}{lcr} \flux\delta_{ij} |\mbhw \rangle, & i,j\leq M \\ 0, & j>M.
                                          \end{array}\right.\label{p2}
\eea
It is also normalized [$(p)_\flux$ denotes the Pochhammer symbol] and invariant under  $\rmu(M)\times \rmu(N-M)\subset \rmu(N)$ transformations. 
\end{prop}
\noindent The proof is left for the Appendix \ref{hwapp}. The vector $|\mbhw \rangle$ is the boson analogue of the fermion state $|\Phi_0\ra$ in  \eqref{GS}. 
The determinant structure of $|\mbhw \rangle$ guarantees that  this state is \emph{antisymmetric} under  electron exchange  
(i.e. under row exchange of $\adhw$) as long as $\flux$ is odd. Otherwise, a  statistical transmutation occurs for the fermion mixture.

Let us identify the ladder operators. It can be seen that any other state $S_{jk}|\mbhw \rangle$ is either zero or has lower weight than $W=[\flux,\stackrel{M}{\dots},\flux,0,\stackrel{N-M}{\dots},0]$. 
Indeed, using the commutation relations \eqref{commurel}, 
\be
[S_{ii},S_{jk}]=\delta_{ij}S_{ik}-\delta_{ik}S_{ji}\Rightarrow S_{ii} S_{jk}|\mbhw \rangle=(W_i+\delta_{ij}-\delta_{ik})S_{jk}|\mbhw \rangle.\label{commuw}
\ee
Actually, from property \eqref{p2}, $S_{jk}|\mbhw \rangle$ gives a non-zero vector of  weight  $w\not=W$ only when $k\leq M<j$.  The resulting vector $S_{jk}|\mbhw \rangle$ has 
the same structure as $|\mbhw \rangle$ but replacing column $k$, $(a^\dag_{k1},\dots, a^\dag_{kM})^t$,  of 
$\adhw$ in $|\mbhw \rangle$ by column $j$, $(a^\dag_{j1},\dots, a^\dag_{jM})^t$ of $A^\dag$ in \eqref{aadag}. When $i\leq M$, the weight component $w_i$ of $S_{jk}|\mbhw \rangle$ is $w_i=W_i-\delta_{ik}=\flux-\delta_{ik}$. 
When $i>M$  the weight component $w_i$ of $S_{jk}|\mbhw \rangle$ is $w_i=0+\delta_{ij}$. Only the weight components $W_j$ and $W_k$ are shifted: $W_j$ increases by 1 and $W_k$ decreases by 1. 
Therefore, $S_{jk}|\mbhw \rangle$ becomes of lower weight since the first non-vanishing coefficient of $W-w$ is $(W-w)_k=1>0$. In this sense, $S_{jk}$, with  $j>k$ acts as  a \emph{lowering} ladder operator; 
It transfers one electron from component $k$ 
into component $j>k$. Of special interest are the step 1 lowering operators $S_{i,i-1}$, from which we can obtain the action of any other lowering operator making use of the 
recursion formulas 
\be
S_{i,i-k}=[S_{i,i-1},S_{i-1,i-k}], k>0. 
\ee
The same argument can be applied to raising ladder operators $S_{kj}$ with $j>k$. We shall provide an explicit expression for the matrix elements of step 1 lowering  $S_{i,i-1}$ and raising  
$S_{i-1,i}$ operators for any IR of $\rmu(N)$ of a given HW in Section  \ref{matsec}. 

Let us see how to label and graphically represent  basis states  of any IR of shape $h$ of $\rmu(N)$  in Young tableau and Gelfand notation. We shall pay special attention to the Hilbert space $\mathcal H_N[L^M]$,

\subsection{Young tableaux, Gelfand and Fock basis states}\label{recyoungel}

Young tableaux are constructed by filling out rows (resp. columns) of the corresponding Young diagram with components $i=1,\dots,N$ in non-decreasing (resp. strictly increasing) order from left to right (resp. from top to bottom). 
For example, for filling factor $M=2$, $\flux=7$ Landau sites and $N=4$ fermion components/flavors, the following Young tableau 
\be
 \young(1112223,2333444)\label{ytsf}
\ee
is in the standard form. The occupancy number  $n_{\mu i}$ described after \eqref{Fockstate} can be calculated as the number of times that the state $i$ appears in the row $\mu$ (counting downwards) of the tableau. 
In the previous example we have
\[n_{11}=3, n_{12}=3, n_{13}=1; n_{22}=1, n_{23}=3, n_{24}=3,
 \]
and zero the rest. It is clear that $\sum_{i=1}^Nn_{\mu i}=\flux, \mu=1,\dots,M$, that is, each electron $\mu=1,\dots,M$ has $\flux$   Landau sites available  (``it carries $\flux$ flux quanta''). 
The highest weight vector $|\mbhw \rangle$   in \eqref{highestweight} is written in Young tableau notation as
\be
\Yvcentermath1  |\mbhw \rangle=\young(1\mydots 1,:::,M\mydots M)\:.
\ee

To subsequently write matrix elements of $\rmu(N)$-spin operators $S_{ij}$ in a compact form (see Section \ref{matsec}), it is convenient to introduce the Gelfand-Tsetlin notation for vectors as 
triangular patterns of non-negative integer numbers $m_{i,j}$ of the form 
\be\label{mVec}
|\mb\ra=\left|\begin{matrix}
 m_{1,N} &&  m_{2,N} && \dots && m_{N-1,N} && m_{N,N}&  \\ 
 & m_{1,N-1} && \dots && \dots && m_{N-1,N-1} & &  \\ & &  \dots && \dots && \dots & & \\
 & && & m_{1,1} & && & & 
\end{matrix}\right\rangle
\ee
obeying the betweenness conditions  
\be 
m_{i,j}\geq m_{i,j-1}\geq m_{i+1,j}\geq 0.\label{betweenness} 
\ee
That is, each number in the pattern $\mb$ is constrained to vary between its two closest upper neighbors. Sometimes we shall denote a Gelfand pattern by its $N$ rows $\mb=\{m_N,\dots,m_1\}$. The relation between a Young tableau 
and the corresponding Gelfand pattern $\mb=\{m_N,\dots,m_1\}$ is built as follows (the prescription  applies to a Young tableau of general shape, not only rectangular $[L^M]$):
\begin{itemize}
 \item The top row $m_N$ is read off the shape of the tableau, and it coincides with the highest weight. In terms of the occupancy numbers $n_{\mu i}$,  we have 
 \begin{equation}\label{mTopRow}
 	m_N=[\sum_{i=1}^N n_{1,i},\stackrel{M}{\dots},\sum_{i=1}^N n_{N,i}]\,.
 \end{equation}
 \item The second row $m_{N-1}$  is read off the shape of the tableau  that remains after all boxes containing the component/flavor $i=N$ are removed, that is, $m_{i,N-1}=m_{i,N}-n_{i,N}$.
\item $\dots$
\item $m_{N-k}$ is read off the shape of the tableau that remains after all boxes containing the flavors $i=N,N-1,\dots,N-k+1$ are removed, that is, $m_{i,N-k}=m_{i,N-k+1}-n_{i,N-k+1}$.
\item $\dots$ 
\item $m_2$    is read off the shape of the tableau that remains after all remaining boxes containing $i=3$ are removed.
\item Finally, $m_1$    is read off the shape of the tableau that remains after all remaining boxes containing $i=2$  are removed.
\end{itemize}

For example, for the Young tableau  \eqref{ytsf} we have 
\be
 \Yvcentermath1 \young(1112223,2333444)= \left|\begin{matrix} 7 &&  7 && 0 && 0 & \\  & 7 && 4 && 0&  &  \\ & & 6 && 1 & & & \\  & & & 3 & & & & \end{matrix}\right\rangle.\label{gytsf}
\ee

Let us work out some particular examples, for the sake of clarity, before  stating more general formulas.

\subsubsection{U(2) quantum Hall ferromagnet at filling factor $M=1$}

Let us describe the simplest example of a QHF where each Landau site  accommodates $M=1$ electron with $N=2$ flavors, for example, a spin $1/2$ electron
\be
\Yvcentermath1\young(1)=|\uparrow\ra, \quad  \young(2)=|\downarrow\ra.\label{spinupdown}
\ee
For $\flux$ Landau sites, the Hilbert space $\mathcal H_2[L^1]$ basis vectors can be labeled by the number $L_1$ of spin-up (flavor $i=1$) electrons in Young tableau, Gelfand and Fock (boson and fermion) forms as
\begin{equation}
	\Yvcentermath1\overbrace{\young(1\mydots12\mydots 2)}^ {L_1+L_2=L}=
	\left|\begin{matrix} 
		\flux &&  0  \\  & L_1 &
	\end{matrix}\right\rangle=
	\frac{(a_{11}^\dag)^{L_1}(a_{21}^\dag)^{L_2}}{\sqrt{L_1!L_2!}}|0\ra_\mathrm{F}=
	\frac{1}{\sqrt{L!}}\sum_{\sigma\in \mathfrak{S}_L}\prod_{\alpha=1}^{L_1}c_1^{\dagger}\big(\sigma(\alpha)\big)\prod_{\beta=L_1+1}^{L_2}c_2^{\dagger}\big(\sigma(\beta)\big)|0\ra_\mathrm{F}\,,\label{fullysym}
\end{equation}
where $\mathfrak{S}_L$ is the symmetric group of degree $L$ and $\sigma$ a permutation. 
Moreover, for this case, a Dicke state representation $\{|j,m\ra, m=-j,\dots,j\}$ is also possible, with total angular momentum $j=L/2$ and spin third component $m=(2L_1-L)/2$. The highest ($L_1=L$) 
and lowest ($L_2=L$)  weight states correspond to angular momentum third components $m=L/2=j$ and $m=-L/2=-j$, respectively. 
The Hilbert space dimension is clearly 
$D[L^1]=L+1=2j+1$.

This is time for a clarification. Even though we are using the equality sign ``$=$'' in \eqref{fullysym}, to be precise, each of the vectors in those equalities belong to different vector spaces. 
That is, they are different mathematical ways of representing the same physical state. 
However, we will keep this little abuse of  notation in the hope that no confusion arises.
 
\subsubsection{U(4) quantum Hall ferromagnet at filling factor $M=2$}\label{bilayersec}

Let us consider now  a bilayer system (with top $t$ and bottom $b$ layers) where each Landau site accommodates $M=2$ electrons with $N=4$ flavors
\be
\Yvcentermath1\young(1)=|\uparrow t\ra,\quad \young(2)=|\uparrow b\ra, \quad\young(3)=|\downarrow t\ra,\quad \young(4)=|\downarrow b\ra.\label{spinppin}
\ee
The basis states of $[L^2]$ are given by the Gelfand vectors and their betweenness conditions
\be 
 |\mb\ra=\left|\begin{matrix}
 \flux &&  \flux && 0 && 0 & \\ 
 & \flux && m_{23} && 0&  &  \\ & &  m_{12} && m_{22} & & & \\
 & & & m_{11} & & & &
\end{matrix}\right\rangle,\quad  \left\{\ba{l} \flux\geq m_{23}\geq 0,\\ \flux\geq m_{12}\geq m_{23},\\ m_{23}\geq m_{22}\geq 0, \\ m_{12}\geq m_{11} \geq m_{22}.     \ea\right.\label{betweenU3}
\ee
In this case, the basis vectors are indexed by four labels $(m_{11},m_{12},m_{22},m_{23})$. Particular examples are the highest- $|\mbhw \ra$ and the lowest-  $|\mblw \ra$ weight states 
\bea |\mbhw \ra&=&\Yvcentermath1 \young(1\mydots 1,2\mydots 2)=\left|\begin{matrix}
 \flux &&  \flux && 0 && 0 & \\ 
 & \flux && \flux && 0&  &  \\ & & \flux && \flux & & & \\
 & & & \flux & & & &
\end{matrix}\right\rangle={[(1)_\flux (2)_\flux]^{-\frac{1}{2}}}{\left|\begin{matrix} a^\dag_{11}  & a^\dag_{21}  \\  a^\dag_{12} & a^\dag_{22}\end{matrix}\right|^{\flux}}|0\ra_\mathrm{F}=
\prod_{\alpha=1}^{L} c_1^\dag(\alpha)c_2^\dag(\alpha)|0\ra_\mathrm{F},\label{highestweightu4}, \\ 
 |\mblw \ra&=&\Yvcentermath1 \young(3\mydots 3,4\mydots 4)=\left|\begin{matrix}
 \flux &&  \flux && 0 && 0 & \\ 
 & \flux && 0 && 0&  &  \\ & & 0 && 0 & & & \\
 & & & 0 & & & &
\end{matrix}\right\rangle={[(1)_\flux (2)_\flux]^{-\frac{1}{2}}}{\left|\begin{matrix} a^\dag_{31}  & a^\dag_{41}  \\  a^\dag_{32} & a^\dag_{42}\end{matrix}\right|^{\flux}}|0\ra_\mathrm{F}=
\prod_{\alpha=1}^{L} c_3^\dag(\alpha)c_4^\dag(\alpha)|0\ra_\mathrm{F},\label{hwlwbilayer}
\eea
in Young tableau, Gelfand and Fock (boson and fermion) notation, respectively. The relation between Gelfand and Fock states for general $\flux$ is a bit more involved for states other than the highest and lowest weight; 
therefore, we leave the general prescriptions for the Appendix \ref{appGelfandFock}. An alternative basis for this case was, noted by 
\be |{}{}_{q_t,q_b}^{j,m}\ra,\; q_t,q_b=-j,\dots,j,\; 0\leq 2j+m\leq\flux,\ee
has been given in  \cite{GrassCSBLQH}, where $j$ (half-integer) represents an angular momentum  and $m$ (integer)  is related to a population  imbalance between layers $t$ and $b$ (both non-negative).

From the betweenness conditions \eqref{betweenU3}, one can easily compute the dimension of the IR $[\flux^2]$ of $\rmu(4)$  as
\be
D[\flux^2]= \sum_{m_{23}=0}^{\flux } \sum_{m_{12}=m_{23}}^{\flux } \sum_{m_{22}=0}^{m_{23}} \sum_{m_{11}=m_{22}}^{m_{12}} 1= \frac{1}{12} (\flux +1) (\flux +2)^2 (\flux +3).\label{dimN4}
\ee
Note that  $D[\flux^2]$ grows as  $\flux^4/12$ for large $\flux$. We shall recover in Section  \ref{gendimsec} the expression \eqref{dimN4}  as a particular case of the  so called ``hook-length'' general formula, 
which is a special case of the Weyl's character formula (see e.g. \cite{Barut}). 

In Appendix \ref{spinpspinapp} we explicitly work out the case $\flux=1$, for which $D[1^2]=6$, thus recovering the dimension $\tbinom{N}{M}$ of the totally antisymmetric IR $[1,1]$ of  $\rmu(4)$. 
The corresponding basis vectors for this case can be divided into two spin/pseudospin (layer) sectors, and we shall make use of them when writing Grassmannian $\mathbb{G}^4_2$ coherent states later in equation \eqref{cohbilayer}.

\subsubsection{U(6) quantum Hall ferromagnet at filling factor $M=3$}\label{trilayersec}

Let us consider now  a trilayer system (with top $t$, central $c$, and bottom $b$ layers) where each Landau site accommodates $M=3$ electrons with $N=6$ flavors
\be \Yvcentermath1\young(1)=|\uparrow t\ra,\quad\young(2)=|\uparrow c\ra,  \quad\young(3)=|\uparrow b\ra,\quad  \young(4)=|\downarrow t\ra,\quad\young(5)=|\downarrow c\ra, \quad\young(6)=|\downarrow b).\ee
The basis states of $[\flux^3]$ are given by the Gelfand vectors  indexed by 9 labels
$$(m_{11}; m_{12},m_{22};m_{13},m_{23},m_{33}; m_{24},m_{34}; m_{35}).$$
In particular, the HW state 
\be
|\mbhw \rangle=\Yvcentermath1 \young(1\mydots 1,2\mydots 2,3\mydots 3)=
{[(1)_\flux(2)_\flux(3)_\flux]^{-1/2}}{\left|\begin{matrix} a^\dag_{11}  & a^\dag_{21} & a^\dag_{31} \\  a^\dag_{12} & a^\dag_{22} & a^\dag_{32} \\  a^\dag_{13} & a^\dag_{23} & a^\dag_{33}\end{matrix}\right|^{\flux}}|0\ra_\mathrm{F}
=\prod_{\alpha=1}^{L} c_1^\dag(\alpha)c_2^\dag(\alpha)c_3^\dag(\alpha)|0\ra_\mathrm{F}
\label{highestweightu6}\ee
corresponds to the Gelfand vector with all 9 labels $m_{ij}=\flux$. As we did in \eqref{dimN4}, from the betweenness conditions \eqref{betweenness} of  these labels, one can compute the dimension 
\be
D[\flux^3]=\frac{(\flux +1) (\flux +2)^2 (\flux +3)^3 (\flux +4)^2 (\flux +5)}{8640}.\label{dimN6}
\ee
For $\flux=1$ we have $D[1^3]=20$, thus recovering the dimension $\tbinom{N}{M}$ of the totally antisymmetric IR $[1^3]$ of $\rmu(6)$. 
Note that $D[\flux^3]$  grows like $\flux^9/8640$ for large $\flux$.

\subsection{General dimension formulas}\label{gendimsec}

The dimension of the carrier Hilbert space of a IR of $\rmu(N)$ with general HW $m_N=[m_{1N},\dots,m_{NN}]$ (a partition of $P$) is given by the Weyl dimension formula (see e.g. \cite{Barut})
\be
D[m_N]=\frac{\prod_{ i<j} (m_{iN}-m_{jN}+j-i)}{\prod_{i=1}^{N-1}i!}.\label{dimform}
\ee
It can also be written with the  so called ``hook formula''  
\be
D[m_N]=\prod_{i,j} \frac{N+j-i}{|h_{m_N}(i,j)|},
\ee
where $|h_{m_N}(i,j)|$ is the length of the hook located at the box/cell position $(i,j)$ (row, column) of the corresponding Young diagram of shape $m_N$. The hook $h_{m_N}(i,j)$ 
is the set of cells/boxes $(k,l)$  such that $k = i$ and $l \geq j$  or $k\geq i$ and $l = j$. The hook length  $|h_{m_N}(i,j)|$ is the number of cells/boxes in $h_{m_N}(i,j)$. 

These formulas correspond to the number of independent Gelfand patterns $\mb$ fulfilling the betweenness conditions \eqref{betweenness}, and also to the number of different 
Young tableau arrangements.

For rectangular Yound diagrams $m_N=[L^{M},0^{N-M}]$ the dimension formula \eqref{dimform}  acquires the form
\be
D{[\flux^M]}=\frac{\prod_{i=N-M+1}^{N}\binom{i+\flux-1}{i-1}}{\prod_{i=2}^M\binom{i+\flux-1}{i-1}},\label{dimensionlambdaM}
\ee
This formula reproduces the previous particular examples. Note that $D{[\flux^M]}=D{[\flux^{N-M}]}$ (conjugated representation).

\section{Matrix elements of U(N)-spin collective operators}\label{matsec}

In this Section we shall provide explicit expressions for matrix elements $\la \mb'|S_{ij}|\mb\ra$ of $\rmu(N)$-spin operators $S_{ij}$  \eqref{Aij} in the Gelfand-Tsetlin basis $\{|\mb\ra\}$. 
We have already given some indications in Section \ref{recyoungel}. In fact, recursion formulas 
\be S_{i,i-l}=[S_{i,i-1},S_{i-1,i-l}]\,,\;  S_{i-l,i}=[S_{i-l,i-1},S_{i-1,i}]\,, \;l>1\,,\label{recurrence}\ee
allow us to obtain any non diagonal operator $S_{ij}$ matrix element from the matrix elements of step 1 lowering  $S_{i,i-1}$ and raising  $S_{i-1,i}$ operators. 
Let us consider an arbitrary  IR of $\rmu(N)$ of HW $m_N$. Denoting by $\bar{m}_k=\sum_{i=1}^k m_{ik}$, $k=1,\dots,N$ the sum of $k$-th row of a pattern $\mb$, and setting $\bar{m}_0\equiv 0$, the action of diagonal operators $S_{kk}$ on 
an arbitrary Gelfand state $|\mb\ra$ is
\be\label{Skk}
S_{kk}|\mb\ra=(\bar{m}_k-\bar{m}_{k-1})|\mb\ra,
\ee
which reproduces the expressions \eqref{p2} for the highest-weight vector $|\mbhw \ra$ with rows  $m_{N-k}=[\flux^{M},0^{N-M-k}]$ for $0\leq k< N-M$ and 
$m_{k}=[\flux^{k}]$ for  $1\leq k\leq M$. The linear Casimir $C_1=\sum_{k=1}^N S_{kk}$ fulfills $C_1|\mb\ra=M\flux|\mb\ra$, the eigenvalue $P=M\flux$ being the total number of particles.

Let us denote by $\mathbb{e}_{jk}$ the ``auxiliary pattern'' with 1 at place $(j,k)$ and zeros elsewhere [we call it ``pattern'' because it has the triangular shape, although it does not necessarily 
fulfill the betweenness conditions \eqref{betweenness}]. The action of step 1 lowering $S_{-k}\equiv S_{k,k-1}$ and rising operators 
$S_{+k}\equiv S_{k-1,k}$ is given by \cite{Barut}
\be\label{S+k}
S_{\pm k}|\mb\ra=\sum_{j=1}^{k-1} c^\pm_{j,k-1}(\mb)|\mb\pm\mathbb{e}_{j,k-1}\ra,
\ee
with coefficients
\be
c_{j,k-1}^\pm(\mb)=\left(-\frac{\prod_{i=1}^N(m'_{ik}-m'_{j,k-1}+\frac{1\mp 1}{2})\prod_{i=1}^{k-2}(m'_{i,k-2}-m'_{j,k-1}-\frac{1\pm 1}{2})}{\prod_{i\not=j}(m'_{i,k-1}-m'_{j,k-1})(m'_{i,k-1}-m'_{j,k-1}\mp 1)}\right)^{1/2},\label{coef}
\ee
where $m'_{ik}=m_{ik}-i$ and $c_{j,k-1}^\pm(\mb)\equiv 0$ whenever any indeterminacy arises. In fact, from the commutation relations \eqref{commuw}, the weight $w'$ of $S_{-k}|\mb\rangle$ is given by
\be
S_{ii}S_{-k}|\mb\rangle=(w_i+\delta_{i,k}-\delta_{i,k-1})S_{-k}|\mb\rangle=w'_iS_{-k}|\mb\rangle,
\ee
and therefore, $S_{-k}|\mb\rangle$ becomes of lower weight than $|\mb\ra$ since the first non-vanishing coefficient of $w-w'$ is $(w-w')_{k-1}=1>0$. From the definition \eqref{coef} one can prove that 
\be
c^\pm_{j,k-1}(\mb)=c^{\mp}_{j,k-1}(\mb\pm\mathbb{e}_{j,k-1}),
\ee
which means that $S_{+k}^\dag=S_{-k}$. Also, applying induction and the recurrence formulas \eqref{recurrence}, we obtain $S_{k,k-l}^\dag=S_{k-l,k}$. Therefore, we can construct proper hermitian $\rmu(N)$-spin operators 
as: $\mathcal{S}_{ii}=S_{ii}$, $\mathcal{S}_{ij}=S_{ij}+S_{ji}$ and $\tilde{\mathcal{S}}_{ij}=\ic(S_{ij}-S_{ji})$, $i<j\leq N$, with $\ic$ the imaginary unit. 
In Appendix \ref{appmatrixelements} we provide explicit expressions of these $\rmu(N)$-spin matrix elements for particularly interesting cases.

For completeness, we shall provide the eigenvalues of the $N$ invariant (Casimir) $\rmu(N)$ operators $C_p$ belonging to the enveloping algebra, whose expression is given by $p$ powers of the operators $S_{ij}$ as
\be C_p=S_{i_1,i_2}S_{i_2i_3}\dots S_{i_{p-1}i_p}S_{i_pi_1}, \; p=1,\dots,N,\label{casimires}\ee
where sum on repeated indices is understood. That is, $C_p$ is of degree $p$. We have already argued that $C_1|\mb\ra=\bar{m}_N|\mb\ra$ with $\bar{m}_N=\sum_{i=1}^N m_{iN}$. For 
$m_N=[\flux^M,0^{N-M}]$, the  $C_1$ eigenvalue is $\bar{m}_N=M\flux=P$ the total number of particles. The eigenvalues of $C_p$ on 
the carrier Hilbert space  of  $m_N$ are given in \cite{Barut} and they are constructed as follows. Let $B$ a $N\times N$ square matrix with entries 
\[b_{ij}=(m_{iN}+N-i)\delta_{ij}-u_{ij},\quad u_{ij}=\left\{\ba{lcr} 1 & \mathrm{for} & i<j\\ 0& \mathrm{for} & i\geq j,\ea\right.\]
and let $J$ be the $N\times N$  all-ones matrix (that is, $J_{ij}=1$). Then the spectrum of the Casimir operators is given by
\be
C_p(m_N)=\tr(B^pJ),\label{casimirval}
\ee
where $B^p$ is the $p$-th power of $B$. The quadratic  Casimir operator $C_2$ plays a fundamental role as the $\rmu(N)$ invariant part of 
the QHF Hamiltonian and we shall pay special attention to it. 
In particular, the eigenvalue of the quadratic Casimir operator is simply given in general by $C_2(m_N)=\sum_{i=1}^N m_{iN}(m_{iN}+N+1-2i)$ and, 
for the case of  $m_N=[\flux^M,0^{N-M}]$, the expression reduces to  $C_2(m_N)=M\flux(\flux+N-M)$.

\section{Grassmannian coherent states and nonlinear sigma models}\label{Grassec}

\subsection{Grassmannian coherent states}

We have seen in Proposition \ref{groundstate} that the HW (ground) state $|\mbhw\ra$ is invariant under the subgroup $\rmu(M)\times \rmu(N-M)$ of 
$\rmu(N)$. Therefore $|\mbhw\ra$ breaks the $\rmu(N)$ symmetry since a general $\rmu(N)$ rotation mixes the first $M$ (``spontaneously chosen'') occupied  internal orbitals with the remainder $(N-M)$ unoccupied ones. 
This structure is also very relevant for systems with particle-hole symmetry, like in nuclear and molecular models \cite{Casten1993,Frankvanisacker,Iachellolevine,iachello_arima_1987}.
$\rmu(N)$-spin-wave excitations occur in  $\rmu(N)$ QHF. These coherent excitations (named ``skyrmions'') turn out to be described by 
a ferromagnetic order parameter associated to this spontaneous symmetry breaking and  labeled by 
$(N-M)\times M$ complex matrices $Z$ parametrizing the  complex Grassmannian 
coset $\mathbb{G}_M^N=\rmu(N)/[\rmu(M)\times \rmu(N-M)]$. This parametrization is related to the Bruhat-Iwasawa block matrix decomposition (see e.g.  Chapter 3 of Ref. \cite{Barut}) 
of the complexification $\mathrm{GL}(N,\mathbb{C})$ of $\rmu(N)$. For 
the fundamental $N$-dimensional representation, this block matrix decomposition reads
\be  U=\left(\ba{c|c}  A & B
\\ \hline C & D \ea\right) =\underbrace{\left(\ba{c|c}  \Delta_1 & - Z^\dag\Delta_2
\\ \hline Z\Delta_1 & \Delta_2 \ea\right)}_{Q(Z)}\underbrace{\left(\ba{c|c} V_1& 0
\\ \hline 0 &V_2\ea\right)}_V\label{cosetrep}
\ee
where $A$ and $D$ are invertible square complex matrices of orders $M$ and $N-M$, respectively, with 
\be Z=CA^{-1}, \quad \Delta_1=(AA^\dag)^{1/2}=(\mathbb{1}_M+ Z^\dag Z)^{-1/2}, \quad \Delta_2=(DD^\dag)^{1/2}=(\mathbb{1}_{N-M}+ Z Z^\dag)^{-1/2}\ee 
and $V_1=\Delta_1^{-1}A\in \rmu(M), V_2=\Delta_2^{-1}D\in \rmu(N-M)$ are unitary matrices. The normalization matrix factors $\Delta_{1,2}$ are related by the Woodbury matrix identity $\Delta_1^2=\mathbb{1}_M-Z^\dag\Delta_2^2 Z$. 
Complex $(N-M)\times M$ matrix points $Z$ on the Grassmann manifold $\mathbb{G}_M^N$ are associated to quotient representatives $Q(Z)\in \rmu(N)$.

Let us firstly discuss the simplest $N=2$ case.  
The decomposition \eqref{cosetrep} for $U\in \rmu(2)$ adopts the form
\be
U=\left(\ba{cc}  a & b
\\ c &d \ea\right)= \underbrace{\left(\ba{cc} \delta  & - \bar{z}\delta
\\ z\delta & \delta\ea\right)}_{Q(z)}\left(\ba{cc} a/\delta  & 0
\\ 0 & d/\delta\ea\right),\quad \delta=(1+|z|^2)^{-1/2},\label{cosetrepu2}
\ee
which is adapted to the quotient $\mathbb{G}_1^2=\rmu(2)/[\rmu(1)\times \rmu(1)]=\mathbb{S}^2$ (the two-sphere), with $z=c/a=\tan(\theta/2)e^{\ic\phi}$ the stereographic projection of a point 
$(\theta, \phi)$ (polar and azimuthal angles)  of the sphere $\mathbb{S}^2$ onto the complex plane $\mathbb{C}\ni z$. Let us consider filling factor $M=1$. 
The $M\times N$ creation operator matrix $A^\dag$ in \eqref{aadag} reduces to $A^\dag=(a^\dag_{11},a^\dag_{21})$. 
Denote by $Q_\mathrm{hw}(z)=(\delta,z\delta)^t$ the first column of $Q(z)$ in  \eqref{cosetrepu2}. 
Coherent excitations above the HW vector  $|\mbhw\ra=\frac{(a^\dag_{11})^\flux}{\sqrt{\flux!}}|0\ra_{\mathrm{F}}$ for filling factor $M=1$ can be written as a two-mode Bose-Einstein condensate of the form
\bea
|z\rangle^L&=&\frac{A^\dag Q_\mathrm{hw}(z)|0\rangle_\mathrm{F}}{\sqrt{\flux!}}=\frac{1}{\sqrt{\flux!}}\left(\frac{a_{11}^\dag+za^\dag_{21}}{\sqrt{1+|z|^2}}\right)^{\flux}|0\ra_\mathrm{F}\nonumber\\ 
&=& \big[\underbrace{\cos(\theta/2)|\uparrow\ra+\sin(\theta/2)e^{\ic \phi}|\downarrow\ra}_{|z\ra}\big]^{\otimes \flux}=|z\ra^{\otimes\flux},\label{u2BE}
\eea
where we are using the notation \eqref{spinupdown} for spin up  and down states  at a Landau site, respectively. That is, the spin $j=L/2$ coherent state $|z\ra^L$ adopts the form of a 
symmetric $L$-qubit state. The direct product structure (not entanglement between lattice sites) ensures the underlying translational invariance. For example, for the particular case of $L=2$ (spin $j=1$) we have
\be |z\ra^{ 2}=\frac{(|\uparrow\ra+z|\downarrow\ra)^{\otimes 2}}{1+|z|^2}=\frac{|\uparrow\uparrow\ra+z(|\uparrow\downarrow\ra+|\downarrow\uparrow\ra)+z^2|\downarrow\downarrow\ra}{1+|z|^2},
\ee
where we identify the spin triplet 
\[ \{|1,1\ra=|\uparrow\uparrow\ra,\quad |1,0\ra=\frac{|\uparrow\downarrow\ra+|\downarrow\uparrow\ra}{\sqrt{2}},\quad |1,-1\ra=|\downarrow\downarrow\ra\}\] 
basis  written in terms of the usual Dicke (total angular momentum) states $|j,m\ra$ (with $m=-j,\dots,j$,  the magnetic quantum number) already discussed 
after \eqref{fullysym}. In general, spin $j=L/2$ coherent states can be written in the Dicke basis as (see e.g. \cite{Perelomov,Gazeaubook} for standard references) 
\be
|z\ra^{ 2j}=(1+|z|^2)^j\sum_{m=-j}^j\sqrt{\binom{2j}{j-m}}z^{j-m}|j,m\ra.
\ee
Even further, coherent states $|z\ra^L$ can also be created by applying a $U(2)$ transformation/rotation on the HW vector
\be
|z\ra^L=e^{yS_{21}-\bar{y}S_{12}}|\mbhw\ra=\frac{e^{zS_{21}}|\mbhw\ra}{\sqrt{1+|z|^2}},\; z=\frac{y}{|y|}\tan|y|,\; y=\frac{\theta}{2}e^{\ic\phi},\label{u2BE2}
\ee
where $S_{21}$ is the spin lowering operator, and the relation between  the complex coordinate $y$ and the stereographic projection coordinate $z$ arises from the application of the Baker-Campbell-Hausdorff-Zassenhaus 
factorization formula to the group $\rmu(2)$ \cite{Perelomov,Gazeaubook}. Note that, in the tensor product representation of $\rmu(2)$, all qubits/spins are rotated ``in unison'' to account for translation/permutation symmetry. Other 
non-symmetric definitions of spin coherent states are possible for the group product  $\rmu(2)\times \dots \times \rmu(2)$, in which every qubit/spin is rotated independently of each other (see e.g. \cite{Sugita2003}). Here we 
restrict ourselves to spin $j=L/2$ (Bloch/atomic) symmetric coherent states  introduced a long time ago by  \cite{Radcliffe} and \cite{Gilmore}.  Haldane used them to study the
semi-classical approximation  of 1-D Heisenberg anti-ferromagnetic spin chains, whose continuum dynamics is described by $\rmsu(2)$  NL$\sigma$Ms  \cite{HaldanePLA93,HaldanePLA93-2,HaldanePLA93-3}.

All these construction can be extended to $\rmu(N)$ QHF at filling factor $M$ as follows. Define  $Q_{\mathrm{hw}}(Z)=\left(\ba{c}  \mathbb{1}_M \\  Z \ea\right) \Delta_1$ as the first $M$ columns of $Q(Z)$ 
in \eqref{cosetrep} and split the $M\times N$ creation operator matrix $A^\dag$ in \eqref{aadag} into a 2-block matrix  $A^\dag=(A_{\mathrm{hw}}^\dag|A_{\mathrm{lw}}^\dag)$, where $A_{\mathrm{hw}}^\dag$ 
makes reference to the first $M$ columns  (HW components) and $A_{\mathrm{lw}}^\dag$ to the last $N-M$ columns (LW components). 
Grassmannian $\mathbb{G}^N_M$ coherent states (see \cite{Gazeaubook} for related fermionic coherent states) are then labeled by the  $(N-M)\times M$ complex matrices $Z$ and have the form of a Bose-Einstein condensate
 \be
|Z\rangle^L=\frac{\det(A^\dag Q_{\mathrm{hw}}(Z))^\flux|0\rangle_\mathrm{F}}{\prod_{p=1}^{M} (p)_\flux^{\um}}=\frac{1}{\prod_{p=1}^{M} (p)_\flux^{\um}}
 \left(\frac{\det(A_\mathrm{hw}^\dag+A^\dag_\mathrm{lw}Z)}{\sqrt{\det(\mathbb{1}_M+{Z}^\dag {Z})}}\right)^\flux|0\rangle_\mathrm{F}.
\label{CScalZ}
\ee
Note that $|Z=0\ra^L$ corresponds to the HW state \eqref{highestweight}. As for spin coherent states in \eqref{u2BE2}, Grassmannian coherent states can also be written as a $\rmu(N)$ 
transformation/rotation of the HW vector as 
\bea
|Z\ra^L&=&\exp\left[\sum_{1\leq j\leq M, M+1\leq i\leq N+M}(Y_{ij}S_{ij}-\bar{Y}_{ij}S_{ji})\right]|\mbhw\ra\label{cohrot}\\
&=&
\frac{\exp\left[\sum_{1\leq j\leq M, M+1\leq i\leq N+M}Z_{ij}S_{ij}\right]|\mbhw\ra}{\sqrt{\det(\mathbb{1}_M+{Z}^\dag {Z})}}, \quad Z=Y (Y^\dag Y)^{-\um}\tan(Y^\dag Y)^{\um},\nonumber
\eea
where the relation between  the $(N-M)\times M$ complex matrices $Y$ and $Z$ [similar to the relation between $y$ and $z$ in \eqref{u2BE2}] now arises from the application of 
the Baker-Campbell-Hausdorff-Zassenhaus factorization formula to the 
group $\rmu(N)$ (see e.g. \cite{Gazeaubook} for related fermionic coherent states).
Note that Grassmannian coherent states $|Z\ra^L$ can be seen as a matrix version/extension of spin coherent states $|z\ra^L$. 

Let us explicitly work out a couple of examples related to the bilayer system ($N=4$)  of Section \ref{bilayersec}. We shall denote the states like in  \eqref{spinppin} and talk about spin $\uparrow,\downarrow$ and 
pseudo-spin or layer $t,b$. For filling factor $M=1$ we have 
\be
|Z\ra^L=\frac{[|1\ra+z_2|2\ra+z_3|3\ra+z_4|4\ra]^{\otimes \flux}}{(1+|z_2|^2+|z_3|^2+|z_4|^2)^{\flux/2}},
\ee
where $Z=(1,z_2,z_3,z_4)^t$ denotes a point on the complex projective space $\mathbb{C}P^{3}=\rmu(4)/\rmu(1)\times \rmu(3)$. For filling factor $M=2$, 
the Hilbert space $\mathcal H_4^\alpha[1^2]$ at each Landau/lattice site $\alpha$ has dimension $\tbinom{4}{2}=6$. 
In Appendix \ref{spinpspinapp} we provide a basis (made of spin triplet $|\mathfrak{S}_{\pm 1,0}\ra$ and pseudo-spin triplet   $|\mathfrak{P}_{\pm 1,0}\ra$ states) for $\mathcal H_4^\alpha[1^2]$ adapted to 
the spin-layer intrinsic structure of this case. Coherent states here adopt the form
\bea
|Z\ra^L&=&\frac{1}{\sqrt{2}}
 \frac{\left|\begin{pmatrix} a^\dag_{11}  & a^\dag_{21}  \\  a^\dag_{12} & a^\dag_{22}\end{pmatrix}+\begin{pmatrix} a^\dag_{31}  & a^\dag_{41}  \\  a^\dag_{32} & a^\dag_{42}\end{pmatrix}
 \begin{pmatrix} z_{11} & z_{12}\\ z_{21} & z_{22}\end{pmatrix}\right|^\flux|0\rangle_\mathrm{F}}{\det(\mathbb{1}_2+{Z}^\dag {Z})^{\flux/2}}\nonumber\\   
&=&\frac{\left[\sqrt{2}|\mathfrak{S}_1\ra+(z_{11}+z_{22})|\mathfrak{S}_0\ra+
	\begin{vmatrix}
		 z_{11} & z_{12}\\
		  z_{21} & z_{22}
	 \end{vmatrix}
	|\mathfrak{S}_{-1}\ra+\sqrt{2}z_{12}|\mathfrak{P}_1\ra+(z_{22}-z_{11})|\mathfrak{P}_0\ra-\sqrt{2}z_{21}|\mathfrak{P}_{-1}\ra\right]^{\otimes\flux}}{\det(\mathbb{1}_2+{Z}^\dag {Z})^{\flux/2}}\,,\label{cohbilayer}
\eea   
with $Z=\begin{pmatrix} z_{11} & z_{12}\\ z_{21} & z_{22}\end{pmatrix}$ a matrix point on the Grassmannian $\mathbb{G}^4_2=\rmu(4)/\rmu(2)^2$. 
$|Z\ra^L$ can also be written in terms of Gelfand vectors $|\mathbb{m}\ra$ as the $\rmu(N)$ rotation  \eqref{cohrot} with $\rmu(N)$-spin operators $S_{ij}$ given by their matrix elements \eqref{coef} 
(see e.g.  Appendix \ref{appmatrixelements} for some particular cases).

Coherent states are sometimes called ``semi-classical'' (they exhibit minimal uncertainty, etc.) and 
they are used as variational states to study the semiclassical and thermodynamic limit, specially in quantum phase transitions \cite{Gilmorebook}. We have used $\rmu(4)$ coherent states, introduced by us in \cite{GrassCSBLQH,JPA48}, to study 
their entanglement properties  \cite{JPCMenredobicapa} and the 
phase diagram of bilayer quantum Hall systems at filling factor $M=2$ in \cite{PhysRevB.95.235302,Calixto_2018}, which turn out to reproduce previous results of Ezawa and collaborators \cite{Ezawabisky,EzawaBook}. 
Let's take a closer look to the role of Grasmannian coherent states to construct semi-classical models of $\rmu(N)$ QHF in terms of NL$\sigma$M.

\subsection{Grassmannian nonlinear sigma models} 

In order to study the semi-classical/thermodynamical  limit $L\to\infty$ of $U(N)$ QHF, one has to replace $\rmu(N)$-spin operators $S_{ij}$ by their coherent state expectation values $\la Z|S_{ij}|Z\ra$. 
Let us adopt a compact notation and denote by ${S}$ the operator matrix with operator matrix  entries $S_{ij}$. The corresponding coherent state expectation value matrix is 
\be \mathcal{S}(Z)\equiv\frac{2}{\flux}\la Z|S-\frac{\flux}{2}\mathbb{1}_N|Z\ra^\flux=\;Q(Z)^\dag E_M Q(Z),\quad  E_M=\mathrm{diag}(1,\stackrel{M}{\dots}1,-1,\stackrel{N-M}{\dots},-1)  ,\label{orderpara}
\ee
where $Q(Z)$ is defined in \eqref{cosetrep}. We have renormalized the matrix operator $S$ by $L$ to define the matrix expectation value $\mathcal{S}$ as a density (intensive quantity), 
with a good thermodynamical limit $L\to\infty$. Moreover, we have shifted Cartan $\rmu(N)$-spin operators 
$S_{ii}$ by $\flux/2$ for convenience. 
The complex $N \times N$ matrix $\mathcal{S}(Z)$ plays the role of a ferromagnetic order parameter  
associated  to the symmetry-breaking ground state. Let us take the continuum limit, that is, small lattice constant $\ell\to 0$ and large number of lattice sites $\flux\to \infty$, so that $\alpha\ell\to x=(x_1,x_2)$ 
are coordinates on the plane and the finite difference $(\mathcal{S}(\alpha+1)-\mathcal{S}(\alpha))/\ell\to \partial_x \mathcal{S}(x)$ becomes the derivative; that is, 
the order parameter $\mathcal{S}$ becomes a matrix field at every point ${x}$ of the plane.  
The low energy physics of the $\rmu(N)$ QH ferromagnet [when considering only nearest-neighbor interactions $\mathcal{J}_{\alpha\beta}=
\mathcal{J}\delta_{\alpha,\beta\pm 1}$ in the exchange energy \eqref{sunqhf}] is then described by a NL$\sigma$M field theory with action
\be
A[Z]=\int dx_0dx_1dx_2\left[\tr(E_M{Q}^\dag \partial_{x_0}{Q})+\mathcal{J}\,\tr(\vec{\nabla}\mathcal{S}\cdot \vec{\nabla} \mathcal{S})\right],\label{action}
\ee
where $\partial_{x_0}\equiv \partial_0$ means partial derivative with respect to time $t=x_0$, $\vec{\nabla}=(\partial_{x_1},\partial_{x_2})\equiv (\partial_1,\partial_2)$ is the gradient 
and $\vec{\nabla}\mathcal{S}\cdot \vec{\nabla} \mathcal{S}$ is the scalar product.  
The first (kinetic) term of the action \eqref{action} is the Berry term (provided by the coherent state representation of the path integral quantization (see e.g. \cite{Sachdev,Arovas} 
for more information about the origin of the Berry term, and \cite{PhysRevA.97.012108} for the application of path-integral quantization to indistinguishable particle systems topologically
confined by a magnetic field). The second term describes the energy cost when the order parameter 
is not uniform. The topological current  
\be
J^\mu=\frac{\ic}{16\pi}\varepsilon^{\mu\nu\lambda} \tr(\mathcal{S}\partial_\nu\mathcal{S}\partial_\lambda\mathcal{S})
\ee
($\varepsilon$ is the Levi-Civita antisymmetric symbol in 1+2 dimensions) leads to the topological (Pontryagin) charge or Skyrmion number 
\be
\mathcal{C}=\int dx_1dx_2 J^0. \label{Pontryagineq}
\ee
See e.g. Ref. \cite{Arovas} for more information.

Note that we do not have $N^2$ real field components for $\mathcal{S}$ but only $2M(N-M)$ corresponding to the $(N-M)\times M$ complex matrix $Z$. 
This is due to the  constraints given by the $N$ values \eqref{casimirval} of the $N$ Casimir operators 
\eqref{casimires}. For example, the linear and quadratic Casimir values say that 
\be
C_1=\sum_{i=1}^N \la Z|S_{ii}|Z\ra=M\flux, \quad C_2=\sum_{i,j=1}^N \la Z|S_{ij}S_{ji}|Z\ra=M\flux(\flux+N-M).
\ee
For large $\flux$,  the leading term for the expectation values of quadratic spin powers is $\la Z|S_{ij}S_{ji}|Z\ra\simeq  \la Z|S_{ij}|Z\ra\la Z|S_{ji}|Z\ra$ (spin fluctuations  are negligible in the classical $\flux\to\infty$ limit). 
For  $N=2$, $M=1$ and $\flux=2j$, the linear $C_1=2j$ and quadratic  $C_2=2j(2j+1)$ Casimir constraints  reproduce the well known sphere equation $\vec{{J}}^2=j(j+1)$ for $J_x=(S_{12}+S_{21})/2$, $J_y=(S_{12}-S_{21})/(2\ic)$  and 
$J_{z}=(S_{11}-S_{22})/2$. 

Since the kinetic (Berry) term involves a single time derivative, half of the Grassmannian fields $Z$ are 
conjugate momenta of the other half (that is, The Grassmannian $\mathbb{G}^N_M$ target space is a phase space), thus expecting $M(N-M)$ independent spin-wave modes. 
Given the relation \eqref{orderpara} between the order parameter $\mathcal{S}$ and the Grassmann matrix $Z$, after a little bit of algebra, the spatial part (potential energy) of the Lagrangian \eqref{action} 
can be written in terms of minimal matrix fields $Z$ as \cite{MACFARLANE1979239,CALIXTO1}
\be
\mathcal{J}\,\tr(\vec{\nabla} \mathcal{S}\cdot \vec{\nabla} \mathcal{S})= \mathcal{J}\, \tr \left( \Delta_2^2\vec{\nabla} Z\,\cdot\, \Delta_1^2  \vec{\nabla}  Z^\dag \right),
\ee
where we have used the expression \eqref{orderpara} of $\mathcal{S}(Z)$ in terms of $Q(Z)$ in \eqref{cosetrep}, together with the Woodbury matrix identity $\Delta_2^2 Z =  Z \Delta_1^2$. 
It would be worth revising the classical limit of $\rmu(N)$ quantum Hall ferromagnets for large $\flux\to\infty$ representations, 
considered long time ago by  \cite{AffleckNPB257,AffleckNPB265,AffleckNPB305,Sachdev,Sachdev2,Arovas} for anti-ferromagnets, in boson/fermion mixture picture exposed in this article. This is work in progress.

\section{Conclusions and outlook}\label{conclusec}


In this article we have presented several group-theoretical tools to study interacting $N$-component fermions on a lattice, like $\rmu(N)$ quantum Hall ferromagnets arising from two-body exchange interactions. 
We have restricted ourselves to the lower energy permutation symmetry sector (according to the Lieb-Mattis theorem \cite{Lieb-Mattis_PR1962}) corresponding to fermion mixtures described by rectangular Young diagrams with $M$ rows (the filling factor) and 
$\flux$ columns (Landau/lattice sites). We have provided orthonormal basis vectors of the corresponding Hilbert space in terms of Youn tableaux, Gelfand-Tsetlin patterns $|\mb\ra$ and boson/fermion Fock states. 
We have written general matrix elements of $\rmu(N)$-spin collective operators $S_{ij}$ in the Gelfand-Tsetlin basis. Several particular examples have been explicitly  worked out to better understand the general expressions, specially 
the case of bilayer $\rmu(4)$ quantum Hall systems at filling factor $M=2$ appearing in the literature \cite{HamEzawa,EzawaBook,PRB60,Schliemann,Ezawabisky,fukudamagnetotransport,KunYang3,PhysRevLett.72.732,PhysRevB.51.5138,PhysRevB.54.11644}
Dimension formulas for these irreducible representations of $\rmu(N)$ have also been provided. 
Special attention has been paid to the highest weight state, which can be associated to the ground state of the system. From this perspective, the ``spontaneously chosen'' 
ground state breaks the original $\rmu(N)$ symmetry and the associated  $\rmu(N)$ ferromagnetic order parameter $\mathcal{S}$ [the expectation value of collective $\rmu(N)$-spin operators $S$ in a Grassmannian coherent state $|Z\ra$] 
describes  coherent state excitations in the classical limit, whose dynamics is governed by a Grassmannian nonlinear sigma model.

Restricting to the dominant permutation symmetry sector is a common practice to reduce the computational complexity when dealing with quantum many-body systems. For example, 
critical and chaotic  quantum systems of $P$, $N$-level/component, identical particles (``quNits'',  a higher dimensional generalization of qubits for $N>2$) 
undergoing a quantum phase transition in the thermodynamic (classical) limit $P\to \infty$, are usually studied by restricting 
to the $\tbinom{P+N-1}{P}$-dimensional (the number of ways of exciting $P$ particles with $N$ levels when order does not matter) totally symmetric 
sector $[P]$ in the $N^P$-dimensional  $P$-fold  tensor product $[1]^{\otimes P}=[P]\oplus[P-1,1]\oplus\dots$ of $N$-dimensional (fundamental) irreducible representations $[1]$ of $\rmu(N)$. 
Replacing $[1]^{\otimes P}$ by $[P]$ then reduces the size of the Hilbert space from $N^P$ to $\tbinom{P+N-1}{P}$, which is a great simplification when $N>1$ and $P$ is large. 
The justification of this restriction is that the ground state of the many-body system always belongs to totally symmetric representation, in accordance with the Lieb-Mattis ordering problem. 
In more physical terms, and for the particular example of the Dicke model of super-radiance \cite{DickePR93}, the assumption that the $P$ atoms of $N=2$-levels  are indistinguishable (bosons) is admissible when the emitters are confined to a cavity 
volume $V\ll \ell^3$ much smaller than the scale of the wavelength $\ell$ of the optical transition. However, the role of mixed permutation symmetry sectors in many body quantum systems should 
not be disregarded at higher energies, and we have already made some steps in \cite{nuestroPRE} for the case of critical $N$-level Lipkin-Meshkov-Glick atom models. 
Entanglement characterization of quantum phases in these systems have also been studied \cite{QIP-2021-Entanglement}. This is also our next step for $\rmu(N)$ quantum Hall ferromagnets.

Finally, concerning physical applications and quantum technological implementations, as we have already mentioned in the introduction,  
the  subject of $\rmsu(N)$ fermions and $\rmsu(N)$ magnetism  has  been recently further fueled in condensed matter physics with exciting advances 
in cooling, trapping and manipulating fermionic alkaline-earth atoms trapped in optical lattices. Also, magnetic Skyrmion materials display a robust topological magnetic structure, being  a candidate for 
the next generation of spintronic memory devices. Multilayer quantum Hall arrangements, bearing larger $\rmu(N)$ symmetries,  also display interesting new physics. Such is the case of superconducting properties of  
twisted bilayer (and trilayer) graphene predicted by \cite{Bistritzer12233} and observed by \cite{Cao2018}. Therefore, this is a highly topical subject, in which we believe this article makes a novel (not standard) 
contribution of a fundamental nature. Further perspectives worth exploring have to do with the interplay between quantum information and quantum topological  phases of matter. Namely, the identification 
of topological order by entanglement entropy (see e.g. \cite{Jiang2012,ZengEntangTPT19}).  Indeed, quantum information concepts can be used to
reformulate and characterize topological order. Some of us have already applied quantum information techniques to the characterization of topological insulator phases of graphene analogues 
\cite{silicene1,silicene2,silicene3,silicene4,silicene5} and fosforene \cite{MRX,IJQC}. Also, Schur basis, like the ones discussed here in terms of Young tableaux, have probed 
to be useful for efficient quNit circuits \cite{PhysRevLett.97.170502}.

\section*{Acknowledgements}

We thank the support of the Spanish MICINN  through the project PGC2018-097831-B-I00 and  Junta de Andaluc\'\i a through the projects SOMM17/6105/UGR, UHU-1262561, FQM-381 and FEDER/UJA-1381026. 
AM thanks the Spanish MIU for the FPU19/06376 predoctoral fellowship. We all thank E. P\'erez-Romero for his collaboration in the early stages of this work.

\appendix

\section{Quantum Hall ferromagnets from exchange interactions}\label{exchangeapp}

Let us briefly remind how the Hamiltonian of a $\rmu(N)$ quantum Hall ferromagnet can be derived from  fundamental microscopic two-body (let us say Coulomb) 
interactions between $N$-component electrons (see also \cite{EzawaBook}). The field theoretical expression of the Hamiltonian for two-body  interactions between $N$-component electrons  in $2$-dimensional space is
\be
H_\mathrm{C}=\frac{1}{2}\sum_{i,j=1}^N\int d^2x d^2x' \Psi^\dag_{i}(x)\Psi^\dag_{j}(x')V(|x-x'|)\Psi_{j}(x')\Psi_{i}(x),
\ee
where $V(|x-x'|)$ is the two-body potential and  $\Psi_{i}(x)$ is the electron field, which can be expanded in terms of a set of 
one-body wave (Wannier) functions $\{\psi_\alpha(x)\}$, 
localized around the lattice/Landau sites  $\alpha=1,\dots,\flux$, with $\flux$ the total number of lattice/landau  sites as
\be
\Psi_{i}(x)=\sum_{\alpha=1}^{\flux} c_{i}(\alpha)\psi_\alpha(x).
\ee
In the case of quantum Hall systems in the Landau gauge $\vec{A}(x)=(Bx_2,0,0)$ ($B$ is the constant perpendicular 
magnetic field and $\vec{A}$ is the vector potential), the one-body functions are canonical (harmonic oscillator) coherent states
\be
\psi_k(x)=\frac{1}{\sqrt{\pi^{1/2}\ell_B}}\exp(ikx_1)\exp\left(-\frac{1}{2\ell_B^2}(x_2+k\ell_B^2)^2\right),
\ee
describing a plane wave propagating in the $x_1$ direction with momentum $\hbar k$ ($\ell_B=\sqrt{\hbar/(eB)}$ denotes the 
magnetic length). The probability of finding the electron at $x_2$ 
has a sharp peak at $x_2=-k\ell_B^2$ and a width $\Delta x_2=\ell_B^2\Delta k$, where $\Delta k=2\pi/\lambda_1$ (here $\lambda_1$ denotes 
the $x_1$-size of the system), because the wave number is quantized as $k_n=2\pi n/\lambda_1$, with $n$ an integer. 
Thus, these states are represented by strips on a rectangular geometry occupying an area of 
$\Delta \mathcal{A}=\lambda_1 \Delta x_2=2\pi\ell_B^2$ and defining a \emph{von Neumann lattice} (see e.g.\cite{EzawaBook}). Therefore, the number of Landau/lattice 
sites enclosed by the system of area $\mathcal{A}=\lambda_1\lambda_2$ is  $\flux=\mathcal{A}/\Delta \mathcal{A}=\mathcal{A}/(2\pi\ell_B^2)$, which coincides with the number of 
magnetic flux quanta penetrating the sample, that is, the ratio of the total magnetic flux $B\mathcal{A}$ 
to the magnetic flux quantum $\phi_0=2\pi\hbar/e$. In the symmetric gauge $\vec{A}(x)=(Bx_2,-Bx_1,0)/2$, the 
``strips in a rectangular geometry'' image is replaced by  ``rings in a disk geometry'', 
where the linear momentum $k$ is replaced by the angular momentum $m$ (see \cite{EzawaBook} for more information).

The coefficients $c_{i}(\alpha)$ $\left(c_{i}^{\dagger}(\alpha)\right)$ denote annihilation (creation) operators of electrons of component $i=1,\dots,N$  at site $\alpha$, fulfilling the usual anticommutation rules
\begin{equation}\label{cAnticommutators}
	\Big\{c_i(\alpha),c_j^{\dagger}(\alpha)\Big\}=\delta_{ij}
	\quad,\quad
	\Big\{c_i(\alpha),c_j(\alpha)\Big\}=
	\Big\{c_i^{\dagger}(\alpha),c_j^{\dagger}(\alpha)\Big\}=
	0
	\quad
	\forall i,j=1,\ldots,N\,.
\end{equation} 
It is also important to emphasize that these operators commute among
different Landau/lattice sites,  that is,
\begin{equation}\label{cCommutators}
	\left[c_i(\alpha),c_j^{\dagger}(\beta)\right]=
	\Big[c_i(\alpha),c_j(\beta)\Big]=
	\left[c_i^{\dagger}(\alpha),c_j^{\dagger}(\beta)\right]=
	0, 
	\quad i,j=1,\ldots,N\,,\quad \alpha\neq\beta.
\end{equation} 
Let us denote by 
\be
V_{\beta\alpha \beta'\alpha'}=\frac{1}{2}\int d^2x d^2x' \bar\psi_\beta(x)\psi_\alpha(x)V(|x-x'|)\bar\psi_{\beta'}(x')\psi_{\alpha'}(x').
\ee
The terms that effectively contribute to the energy are $\mathcal{U}_{\beta\beta'}=V_{\beta\beta\beta'\beta'}$ and $\mathcal{J}_{\beta \alpha}=V_{\beta \alpha\alpha\beta}$, corresponding to the 
\emph{direct} (D) and \emph{exchange} (E) energies 
\be
 H_\mathrm{C}= H_\mathrm{C}^\mathrm{D}+ H_\mathrm{C}^\mathrm{E}=\sum_{\beta,\beta'=1}^{\flux} \mathcal{U}_{\beta\beta'}\rho(\beta)\rho(\beta')+
\sum_{\beta,\alpha=1}^{\flux}\sum_{i,j=1}^N \mathcal{J}_{\beta\alpha} c^\dag_{i}(\beta)c^\dag_{j}(\alpha)
c_{j}(\beta)c_{i}(\alpha),
\ee
where $\rho(\beta)=\sum_{i=1}^N c^\dag_{i}(\beta)c_{i}(\beta)$ is the electron number operator at site $\beta$.  In the case of Coulomb interaction, 
the direct term $H_\mathrm{C}^\mathrm{D}$ represents the usual Coulomb energy between two charge distributions $|\psi_\beta(x)|^2$ and $|\psi_{\beta'}(x')|^2$ localized around the Landau/lattice 
sites $\beta$ and $\beta'$, respectively. The exchange term $H_\mathrm{C}^\mathrm{E}$ 
has no classical counterpart and owes its origin to the Pauli exclusion principle. Note that $\mathcal{J}_{\beta\alpha}$ vanishes when there is no overlap 
between the wave functions $\psi_\beta$ and $\psi_\alpha$ at sites $\beta$ and $\alpha$ (for example, for distant sites). One can 
define the $\rmu(N)$-spin operators  at site $\alpha$ by 
\be
{S}_{ij}(\alpha)= c^\dag_{i}(\alpha)c_{j}(\alpha).\label{collectiveS}
\ee
They allow one to write the exchange energy as a  generalized Heisenberg  spin-spin interaction 
\be
H_\mathrm{C}^\mathrm{E}=-\sum_{\alpha,\beta=1}^{\flux}\mathcal{J}_{\alpha\beta}\sum_{i,j=1}^N {S}_{ij}(\alpha) {S}_{ji}(\beta)+N\sum_{\alpha=1}^{\flux}\mathcal{J}_{\alpha\alpha}\rho(\alpha).\label{sunqhf}
\ee 
which depends on the relative $\rmu(N)$-spin orientation at neighboring sites  $\alpha$ and $\beta$. From here the name of ``$\rmu(N)$ quantum Hall ferromagnet'', where all $\rmu(N)$-spins tend to be  equally 
polarized (for $\mathcal{J}_{\alpha\beta}>0$)
to lower the exchange energy $H_\mathrm{C}^\mathrm{E}$. This Hamiltonian is $\rmu(N)$-invariant and therefore the 
$\rmu(N)$-spin ``direction'' is spontaneously chosen. This invariance can be explicitly broken by adding Zeeman, pseudo-Zeeman, layer bias, etc. external couplings \cite{EzawaBook}.

\section{Proof of Proposition \ref{rectdomin}}\label{rectdominproof}

We shall proceed by induction in $\flux$. The $\flux=2$-fold tensor product representation of $\rmu(N)$  decomposes as
\be
\Yvcentermath1  \young(\quad,:,\quad)  \:\otimes\: \young(\quad,:,\quad)\:=  [1^M]\otimes [1^M]=\bigoplus_{k=0}^{k^*} [2^{M-k},1^{2k}]
\quad,\quad k^*=
\begin{cases}
	N-M&\forall M>\lfloor \tfrac{N}{2}\rfloor\\
	M&\forall M\leq\lfloor \tfrac{N}{2}\rfloor
\end{cases}\:,
\label{tensorprod2}
\ee
where we understand $[a^0]=[0]$ for all $a\in\mathbb{N}$. It is clear that $[2^M]\succeq  [2^{M-k},1^{2k}]$ for all $k=0,\dots,k^*$, in accordance with the dominance order definition \eqref{DominanceIneq}. 
Now suppose that $[\flux^M]$ dominates over all Young diagrams arising in $[1^M]^{\otimes\flux}$.  Then, we have to prove by induction 
that $[(\flux+1)^M]$ dominates over all Young diagrams arising in $[1^M]^{\otimes\flux+1}$.\\
Firstly, we shall state an auxiliary lemma.
\begin{lem}\label{LemmaYoung}
	Let $[h]=[h_1,h_2,\ldots,h_N]$ be any Young diagram of $\rmu(N)$. The tensor product $[h]\otimes[1^M]$ between $[h]$ and the totally antisymmetric IR $[1^M]$ leads to a decomposition into Young diagrams with  shape
	\begin{equation}
		[\tilde{h}]=[\tilde{h}_1,\tilde{h}_1,\ldots,\tilde{h}_N]=[h_1+n_1,h_2+n_2,\ldots,h_N+n_N],\quad n_i\in\{0,1\},\quad \sum_{i=1}^{N}n_i=M,\quad \tilde{h}_1\geq\dots\geq\tilde{h}_N.
	\end{equation} 
\end{lem}
\begin{proof}
	It is straightforward taking into consideration the multiplication rules of Young diagrams (section 9.5.1 of \cite{CVI}). Specially the one which states: Reading the resulting diagrams from right to left and starting with the top row, at any point must the number of $a_i$'s encountered  exceed the number of previously encountered
	$a_{i-1}$'s. If we are multiplying any diagram with the totally antisymmetric (one column), every $a_i$ will appear only once  in the new diagrams. Therefore, the aforementioned rule will limit by one the number of boxes per row that we can add to the original diagram to construct the new ones. For instance, using a $\rmu(6)$ diagram and the 
	antisymmetric IR $[1^5]$,
	\begin{equation}
		\begin{gathered}
			\begin{ytableau}
				~&~&~&~\\ 
				~&~&~\\
				~&~\\
				~&~\\ 
			\end{ytableau}
		\end{gathered}
		\:\:\otimes\:\:
		\begin{gathered}
			\begin{ytableau}
				a_1\\
				a_2\\
				a_3\\
				a_4\\
				a_5
			\end{ytableau}
		\end{gathered}\:\:\cong\:\:
		\begin{gathered}
			\begin{ytableau}
				~&~&~&~&a_1\\ 
				~&~&~&a_2\\
				~&~&a_3\\
				~&~&a_4\\
				a_5\\ 
			\end{ytableau}
		\end{gathered}
		\:\:\oplus\:\:
		\begin{gathered}
			\begin{ytableau}
				~&~&~&~&a_1\\ 
				~&~&~&a_2\\
				~&~&a_3\\
				~&~\\
				a_4\\ 
				a_5
			\end{ytableau}
		\end{gathered}
		\:\:\oplus\:\:
		\xcancel{
		\begin{gathered}
			\begin{ytableau}
				~&~&~&~&a_1\\ 
				~&~&~&a_2\\
				~&~&a_3&a_4\\
				~&~&a_5\\ 
			\end{ytableau}
		\end{gathered}
		}
		\:\:\oplus\:\:
		\ldots
	\end{equation}
\end{proof}

It is convenient to name the rectangular diagram as $[h]=[h_1,h_2,\ldots,h_N]=[\flux^M]=[\flux,\overset{M}{\ldots},\flux,0,\overset{N-M}{\ldots},0]$, 
and all Young diagrams arising from  $[1^M]^{\otimes\flux}$ as $[h']=[h'_1,h'_2,\ldots,h'_N]$, including the rectangular one ($h'=h$). Therefore, the dominance of $[\flux^M]$ is written as $[h]\succeq[h']$, or equivalently \eqref{DominanceIneq}, 
\begin{equation}\label{hIneq}
	k\flux=h_1+\dots+h_k\geq h_1'+\dots+h_k' \quad \forall k\in[1,N]\,.
\end{equation}
According to the lemma \ref{LemmaYoung}, the tensor product $[1^M]^{\otimes(\flux+1)}=[1^M]^{\otimes\flux}\otimes[1^M]$ generates the diagrams 
$[\tilde{h}']=[\tilde{h}'_1,\tilde{h}'_2,\ldots,\tilde{h}'_N]=[h'_1+n_1,h'_2+n_2,\ldots,h'_N+n_N]$ with $n_i\in\{0,1\}$ and $\sum_{i=1}^{N}n_i=M$. Among them, there is a new rectangular diagram $[\tilde{h}]=[\tilde{h}_1,\ldots,\tilde{h}_N]=[(\flux+1)^M]=[h_1+1,\ldots,h_M+1,h_{M+1},\ldots,h_N]$, with $h_{M+1},\ldots,h_N=0$. The restriction $\sum_{i=1}^{N}n_i=M$ implies $\sum_{i=1}^{k}n_i\leq k$, which leads to
\begin{equation}
	\tilde{h}_1+\ldots+\tilde{h}_k=h_1+\ldots+h_k+k\geq h_1+\ldots+h_k+n_1+\ldots+n_k \quad \forall k\in[1,N]\,,
\end{equation}
and using the equation \eqref{hIneq},
\begin{equation}
	\tilde{h}_1+\ldots+\tilde{h}_k\geq h_1+\ldots+h_k+n_1+\ldots+n_k\geq h'_1+\ldots+h'_k+n_1+\ldots+n_k=\tilde{h}'_1+\ldots+\tilde{h}'_k \quad \forall k\in[1,N]\,.
\end{equation}
Therefore, considering the dominance order definition, we arrive to $[\tilde{h}]\succeq[\tilde{h}']$, eventually proving that the rectangular Young diagram $[\tilde{h}]=[(\flux+1)^M]$ dominates the other diagrams arising 
from $[1^M]^{\otimes(\flux+1)}$ and concluding the proof by induction.\\ 
 \quelle{$\blacksquare$}

\section{Proof of Proposition \ref{groundstate}}\label{hwapp}

Looking at the structure of 
\[\det(\adhw)=\sum_{\sigma\in \mathfrak{S}_{M}}\mathrm{sgn}(\sigma)\prod_{i=1}^{M} a_{i,\sigma_i}^\dag=\sum_{\mu_1,\dots,\mu_M=1}^{M}\varepsilon_{\mu_1,\dots,\mu_{M}}\prod_{i=1}^{M} a_{i,\mu_i}^\dag,\]
[where $\mathfrak{S}_{M}$ is the symmetric group of degree $M$ and $\varepsilon$ is the Levi-Civita symbol] it is clear that $\det(\adhw)^{\flux}|0\rangle_\mathrm{F}$ is made of $P=M\flux$ particles, as desired. 
The basic boson commutation relations $[ a, a^\dag]=1$ imply that $[ a,\psi( a^\dag)]=\psi'( a^\dag)$ or $ a\psi( a^\dag)|0\ra_\mathrm{F}=\psi'( a^\dag)|0\ra_\mathrm{F}$, where $\psi$ is a function and 
$\psi'$ denotes the formal derivative with respect to the argument. Therefore, let us simply write $a_{i\mu}=\partial/\partial a^\dag_{i\mu}$. In order to prove \eqref{p1}, we have that
\be
\Lambda_{\mu\nu}\det(\adhw)^{\flux }|0\rangle_\mathrm{F}=
\sum_{i=1}^{N}a^\dag_{i\mu}\frac{\partial}{\partial a^\dag_{i\nu}}\det( \adhw)^{\flux }|0\rangle_\mathrm{F}
= \flux \det( \adhw)^{\flux -1}\sum_{i=1}^{M}a^\dag_{i\mu}\frac{\partial}{\partial a^\dag_{i\nu}}\det( \adhw)|0\rangle_\mathrm{F}.\nonumber
\ee
The last summation consists of replacing row $\nu$ by row $\mu$ inside the determinant  $\det( \adhw)$, and therefore we have 
\[\sum_{i=1}^{M}a^\dag_{i\mu}\frac{\partial}{\partial a^\dag_{i\nu}}\det( \adhw)=\delta_{\mu\nu}\det( \adhw),\]
which proves the constraint \eqref{p1}. To prove \eqref{p2}, we follow the same steps as for \eqref{p1}, that is
\be
S_{ij}\det(\adhw)^{\flux }|0\rangle_\mathrm{F}=
\sum_{\mu=1}^{M}a^\dag_{i\mu}\frac{\partial}{\partial a^\dag_{j\mu}}\det( \adhw)^{\flux }|0\rangle_\mathrm{F}
= \flux \det( \adhw)^{\flux -1}\sum_{\mu=1}^{M}a^\dag_{i\mu}\frac{\partial}{\partial a^\dag_{j\mu}}\det( \adhw)|0\rangle_\mathrm{F}.\nonumber
\ee
If $i,j\leq M$, the last summation consists of replacing column $j$ by column $i$ inside the determinant  $\det( \adhw)$, and therefore $S_{ij}|\mbhw \rangle=\flux\delta_{ij} |\mbhw \rangle$, which means 
that  $|\mbhw \rangle$ is invariant under the subgroup $\rmu(M)\subset \rmu(N)$. 
If $j>M$, then column $j$ is absent from  $\det(\adhw)$ and $S_{ij}|\mbhw \rangle=0$, which means that $|\mbhw \rangle$ is in fact invariant under the subgroup $\rmu(M)\times \rmu(N-M)\subset \rmu(N)$. 
Note the similarities with invariance properties of the ground state $|\Phi_0\ra$ of Eq. \eqref{GS}. This will be 
an important fact when discussing the Grassmannian structure associated to $\rmu(N)$ quantum Hall ferromagnets at filling factor $M$ later in Section \ref{Grassec}. 
The other possibilities for $S_{ij}$ correspond to rising and lowering operators and will be discussed later.

It remains to prove that the squared norm of $\det(\adhw)^{\flux }|0\rangle_\mathrm{F}$ is given by $\mathcal{N}_{\flux}=\prod_{p=1}^{M} (p)_\flux$ in \eqref{highestweight}, where $(p)_\flux$ is the 
usual Pochhammer symbol. We proceed by mathematical induction. Firstly we prove that $\mathcal{N}_{1}=M!$. Indeed, \[\langle 0|\det( A_{\mathrm{hw}})\det( \adhw)|0\rangle_\mathrm{F}=\sum_{\sigma\in \mathfrak{S}_{M}} 1=M!.\]
Now we assume that $\langle 0|\det( A_{\mathrm{hw}})^\flux \det( \adhw)^\flux |0\rangle_\mathrm{F}=\mathcal{N}_{\flux}$ and we shall prove that 
\[\langle 0|\det( A_{\mathrm{hw}})^{\flux +1}\det( \adhw)^{\flux +1}|0\rangle_\mathrm{F}=\mathcal{N}_{\flux +1}.\]
Indeed, it can be shown that 
\[\langle 0|\det( A_{\mathrm{hw}})^{\flux +1}\det( \adhw)^{\flux +1}|0\rangle_\mathrm{F}=(\flux +1)_{M}\langle 0|\det( A_{\mathrm{hw}})^{\flux }\det( \adhw)^{\flux }|0\rangle_\mathrm{F}.\]
The proof is cumbersome in general and we shall restrict ourselves to the more maneuverable $M=2$ case, which grasps the essence of the general case. In fact,
\bea \det( A_{\mathrm{hw}})\det( \adhw)^{\flux +1}|0\rangle_\mathrm{F}&=&
\left(\frac{\partial}{\partial a^\dag_{11}}\frac{\partial}{\partial a^\dag_{22}}-\frac{\partial}{\partial a^\dag_{12}}\frac{\partial}{\partial a^\dag_{21}}\right)
(a^\dag_{11}a^\dag_{22}-a^\dag_{12}a^\dag_{21})^{\flux +1}|0\rangle_\mathrm{F}\nonumber\\ &=&(\flux +1)\flux \det( \adhw)^\flux |0\rangle_\mathrm{F}+2(\flux +1)\det( \adhw)^\flux |0\rangle_\mathrm{F}\nonumber\\ 
&=&(\flux +1)_2\det( \adhw)^{\flux }|0\rangle_\mathrm{F}.\nonumber
\eea
In general 
\[\det( A_{\mathrm{hw}})\det( \adhw)^{\flux +1}|0\rangle_\mathrm{F}=(\flux +1)_{M}\det( \adhw)^{\flux }|0\rangle_\mathrm{F}.\] 
To finish, we realize that  $(\flux +1)_{M}\mathcal{N}_{\flux }=\mathcal{N}_{\flux +1}$, which concludes the proof by induction.\\ \quelle{$\blacksquare$}

\section{Relation between Gelfand-Tsetlin and Fock states}\label{appGelfandFock}

We already know the general expression of the HW state $|\mbhw \ra$ in Fock space, given by \eqref{highestweight}. In this expression, the leading principal minor $\det(\adhw)$ of order $M$  of $A^\dag$ plays a fundamental role. 
Remember that the $M\times M$ square submatrix $\adhw$ was obtained from $A^\dag$ in \eqref{aadag} by deleting the last $N-M$ columns. In the proof of Proposition \ref{groundstate}, in the Appendix \ref{hwapp}, we argued that ladder operators 
$S_{ij}, i\not= j$ replace column $j$ by column $i$ inside the minor $\det(\adhw)$. In general, we can obtain $\binom{N}{M}$ different minors of size $M\times M$, corresponding to the different ways one can choose 
$M$ columns from the $N$ columns of $A^\dag$. Let $I=\{i_1,\dots i_M\}$, with $i_\mu<i_{\mu+1}$ (increasing order),  denote one of these $\binom{N}{M}$ column choices and  
\be
A^\dag_I=\begin{pmatrix} a^\dag_{i_1 1}  &\dots & a^\dag_{i_M 1}  \\  \vdots  & 
& \vdots  \\ a^\dag_{i_1 M} &  \dots & a^\dag_{i_M M}\end{pmatrix}
\ee
the corresponding $M\times M$ submatrix of $A^\dag$. The cases $I=\{1,\dots,M\}$ (first $M$ columns) and $I'=\{N-M,\dots,N\}$ (last $M$ columns) are special, since they are related to the 
highest- and lowest-weight states, respectively; actually, we are denoting $A^\dag_{\{1,\dots,M\}}$ simply by $\adhw$.  There are several ways of attaching $n_{\mu,i}$ flux quanta to the electron $\mu\leq M$ with flavor $i\leq N$. 
For a given $I=\{i_1,\dots i_M\}$ containing $i$, let us denote $\{l_I\geq 0\}$ a composition (a partition where order matters) of $n_{\mu,i}$ in the sense that
\be
n_{\mu,i}=\sum_{i_1<\dots<i_\mu<\dots<i_M} l_{\{i_1,\dots, i_M\}},\quad 0\leq i_k\leq N.\label{partn}
\ee
[$i_\mu$ means that we put $i$ in the $\mu$-th place]. Namely, for the example \eqref{ytsf}, we have $M=2$ ($\mu=1, 2$) and we can arrange these compositions into planar tables, 
where sum on column $i$ gives $n_{1i}$ and sum on row $i$ gives $n_{2i}$, 
as follows
\be
\begin{array}{l|ccc} & n_{11}=3 & n_{12}=3 & n_{13}=1 \\ \hline  n_{24}=3 & l_{14} & l_{24}& l_{34} \\ n_{23}=3 & l_{13}& l_{23}& \\ n_{22}=1 & l_{12} & & \end{array}=
\begin{array}{ccc} & &  \\ 1 &1 & 1 \\ 1 &2 & \\  1& & \end{array}+
\begin{array}{ccc} & & \\0 & 2& 1 \\ 2 & 1& \\ 1 & & \end{array} + 
\begin{array}{ccc} & & \\2 & 0& 1 \\ 0 &3 & \\ 1 & & \end{array}.\label{partl}
\ee
Denoting $\Delta_{I}=\det(A^\dag_I)$ a minor of size $M\times M$ of $A^\dag$, a  Gelfand state  $|\mb\ra$ corresponds to the following (un-normalized) Fock state
\be
|\mb\ra\propto \sum_{l(\mb)}\prod_{I=1}^{\binom{N}{M}} \Delta_I^{l_I}|0\ra_\mathrm{F},\label{minorsprod}
\ee
where the sum is extended to all components $l(\mb)$ associated to  $\mb$ (or equivalently, to the occupancy numbers $n_{\mu i}$). Note that $\prod_{I} \Delta_I^{l_I}$ is a homogeneous polynomial of degree $M\flux$ 
in the creation operators $a^\dag_{i\mu}$. For example, taking into account the three components \eqref{partl} of the Gelfand state  \eqref{gytsf} for filling factor $M=2$ and $N=4$ flavors, the corresponding Fock state can be written as
\be
\left(
\Delta_{12}^{1}\Delta_{13}^{1}\Delta_{14}^{1}\Delta_{23}^{2}\Delta_{24}^{1}\Delta_{34}^{1}+
\Delta_{12}^{1}\Delta_{13}^{2}\Delta_{14}^{0}\Delta_{23}^{1}\Delta_{24}^{2}\Delta_{34}^{1}+
\Delta_{12}^{1}\Delta_{13}^{0}\Delta_{14}^{2}\Delta_{23}^{3}\Delta_{24}^{0}\Delta_{34}^{1}\right)
|0\ra_\mathrm{F}.
\ee
This expression gets simpler for highest and lowest weight states. For example, in  \eqref{highestweightu4} and \eqref{highestweightu6} we have seen that the corresponding HW states for $M=2$ and $M=3$ are 
just given in terms of $\Delta_{12}$ and $\Delta_{123}$ (just one single component), respectively. In the same way, the lowest-weight state  $|\mb_{\mathrm{lw}}\ra$ for $N=4$ and $M=2$ is given in terms of only $\Delta_{34}$ in \eqref{hwlwbilayer}.
The computation of compositions \eqref{partn} of the occupancy numbers $n_{\mu i}$ for $M>2$ gets more and more involved since the planar picture \eqref{partl} becomes a higher-dimensional arrangement.

\section{Single Landau site Hilbert space basis for a bilayer $\rmu(4)$ QHF at $M=2$}\label{spinpspinapp}

Let us explicitly work out the case $N=4$, $M=2$ and $\flux=1$, for which the Hilbert space $\mathcal H_4[1^2]$ has  dimension $\tbinom{N}{M}=6$. The corresponding basis vectors for this case can be divided into two sectors: 
the spin-triplet pseudospin-singlet sector
\begin{align}
	|\mathfrak{S}_1\ra=&\,c^\dag_{\uparrow t}c^\dag_{\uparrow b}|0\ra_\mathrm{F}=
	\Yvcentermath1\young(1,2)=
	\left|
	\begin{matrix}
		1 &&  1 && 0 && 0 & \\ 
		& 1 && 1 && 0 &&  \\
		&& 1 && 1 &&& \\
		&&& 1 &&&&
	\end{matrix}
	\right\rangle
	=\frac{1}{\sqrt{2}}
	\left|
	\begin{matrix}
		 a^\dag_{11}  & a^\dag_{21}  \\
		   a^\dag_{12} & a^\dag_{22}
	\end{matrix}
	\right|
	|0\ra_\mathrm{F}\,,\\
	|\mathfrak{S}_{-1}\ra=&\,c^\dag_{\downarrow t}c^\dag_{\downarrow b}|0\ra_\mathrm{F}=
	\Yvcentermath1\young(3,4)=
	\left|
	\begin{matrix}
		1 &&  1 && 0 && 0 & \\ 
		& 1 && 0 && 0 &&  \\
		&& 0 && 0 &&& \\
		&&& 0 &&&&
	\end{matrix}
	\right\rangle
	=\frac{1}{\sqrt{2}}
	\left|
	\begin{matrix}
		a^\dag_{31}  & a^\dag_{41}  \\
		  a^\dag_{32} & a^\dag_{42}
	\end{matrix}
	\right|
	|0\ra_\mathrm{F}\,,\nonumber\\
	|\mathfrak{S}_0\ra=&\,\frac{c^\dag_{\uparrow t}c^\dag_{\downarrow b}+c^\dag_{\downarrow t}c^\dag_{\uparrow b}}{\sqrt{2}}|0\ra_\mathrm{F}=
	\frac{1}{\sqrt{2}}\left(\,\Yvcentermath1\young(1,4)-\Yvcentermath1\young(2,3)\,\right)=
	\frac{1}{\sqrt{2}}
	\left(\,
	\left|
	\begin{matrix}
		1 &&  1 && 0 && 0 & \\ 
		& 1 && 0 && 0 &&  \\
		&& 1 && 0 &&& \\
		&&& 1 &&&&
	\end{matrix}
	\right\rangle
	-
	\left|
	\begin{matrix}
		1 &&  1 && 0 && 0 & \\ 
		& 1 && 1 && 0 &&  \\
		&& 1 && 0 &&& \\
		&&& 0 &&&&
	\end{matrix}
	\right\rangle
	\,\right)
	\nonumber\\
	=&\,\frac{1}{\sqrt{2}}
	\left(\,
	\left|
	\begin{matrix}
		a^\dag_{11}  & a^\dag_{41}  \\
		a^\dag_{12} & a^\dag_{42}
	\end{matrix}
	\right|
	-
	\left|
	\begin{matrix}
		a^\dag_{21}  & a^\dag_{31}  \\
		a^\dag_{22} & a^\dag_{32}
	\end{matrix}
	\right|
	\,\right)
	|0\ra_\mathrm{F}\,.\nonumber
\end{align}
and the pseudospin-triplet spin-singlet sector
\begin{align}
	|\mathfrak{P}_1\ra=&\,c^\dag_{\uparrow t}c^\dag_{\downarrow t}|0\ra_\mathrm{F}=
	\Yvcentermath1\young(1,3)=
	\left|
	\begin{matrix}
		1 &&  1 && 0 && 0 & \\ 
		& 1 && 1 && 0&  &  \\
		 & & 1 && 0 & & & \\
		& & & 1 & & & &
	\end{matrix}
	\right\rangle
	=\frac{1}{\sqrt{2}}
	\left|
	\begin{matrix} 
		a^\dag_{11}  & a^\dag_{31}  \\  a^\dag_{12} & a^\dag_{32}
	\end{matrix}
	\right|
	|0\ra_\mathrm{F},\\
	|\mathfrak{P}_{-1}\ra=&\,c^\dag_{\uparrow b}c^\dag_{\downarrow b}|0\ra_\mathrm{F}=
	\Yvcentermath1\young(2,4)=
	\left|
	\begin{matrix}
		1 &&  1 && 0 && 0 & \\ 
		& 1 && 0 && 0 &&  \\
		&& 1 && 0 &&& \\
		&&& 0 &&&&
	\end{matrix}
	\right\rangle
	=\frac{1}{\sqrt{2}}
	\left|
	\begin{matrix}
		a^\dag_{21}  & a^\dag_{41}  \\
		a^\dag_{22} & a^\dag_{42}
	\end{matrix}
	\right|
	|0\ra_\mathrm{F}\,,\nonumber\\
	|\mathfrak{P}_0\ra=&\,\frac{c^\dag_{\uparrow t}c^\dag_{\downarrow b}-c^\dag_{\downarrow t}c^\dag_{\uparrow b}}{\sqrt{2}}|0\ra_\mathrm{F}=
	\frac{1}{\sqrt{2}}\left(\,\Yvcentermath1\young(1,4)+\Yvcentermath1\young(2,3)\,\right)=
	\frac{1}{\sqrt{2}}
	\left(\,
	\left|
	\begin{matrix}
		1 &&  1 && 0 && 0 & \\ 
		& 1 && 0 && 0 &&  \\
		&& 1 && 0 &&& \\
		&&& 1 &&&&
	\end{matrix}
	\right\rangle
	+
	\left|
	\begin{matrix}
		1 &&  1 && 0 && 0 & \\ 
		& 1 && 1 && 0 &&  \\
		&& 1 && 0 &&& \\
		&&& 0 &&&&
	\end{matrix}
	\right\rangle
	\,\right)
	\nonumber\\
	=&\,\frac{1}{\sqrt{2}}
	\left(\,
	\left|
	\begin{matrix}
		a^\dag_{11}  & a^\dag_{41}  \\
		a^\dag_{12} & a^\dag_{42}
	\end{matrix}
	\right|
	+
	\left|
	\begin{matrix}
		a^\dag_{21}  & a^\dag_{31}  \\
		a^\dag_{22} & a^\dag_{32}
	\end{matrix}
	\right|
	\,\right)
	|0\ra_\mathrm{F}\,.\nonumber
\end{align}

\section{Explicit particular  expressions of $\rmu(N)$-spin matrix elements}\label{appmatrixelements}

In Section \ref{matsec} we have given general formulas \eqref{coef} for the matrix coefficients  $\la\mb'|S_{ij}|\mb\ra$ of the $\rmu(N)$-spin operators $S_{ij}$ in the Gelfand basis $\{|\mb\ra\}$ for a given IR of $\rmu(N)$
characterized by the highest-weight $h=m_N=[m_{1,N},\ldots,m_{N,N}]$ [top row of the  Gelfand vector $|\mb\ra$ in \eqref{mVec}].  In this Appendix we provide particular examples to explain the underlying algorithm. Firstly, we need 
to order the Gelfand basis $\{|\mb\ra\}$. For it, we choose an increasing order of the components $m_{ij}, j<N$ of the Gelfand vector array \eqref{mVec} from top to bottom and from left to right, 
fulfilling the betweenness conditions \eqref{betweenness}.  Schematically, the vectors of the basis are ordered within a list generated by nested indexes
\begin{equation}
	\{\{\ldots\{\{\ldots\{
	\:|\mb\ra\,
	\}_{m_{11}=m_{22}}^{m_{12}}\,
	\ldots\,
	\}_{m_{1,N-2}=m_{2,N-1}}^{m_{1,N-1}}\,
	\}_{m_{N-1,N-1}=m_{N-1,N}}^{m_{N,N}}\,
	\ldots\,
	\}_{m_{2,N-1}=m_{3.N}}^{m_{2,N}}\,
	\}_{m_{1,N-1}=m_{2,N}}^{m_{1,N}}\:.
\end{equation}
For example, this ordering convention coincides with the order in which we add in equation \eqref{dimN4} the basis vectors of the representation $m_4=[\flux^2]$ of $\rmu(4)$ to compute its dimension. 
Note that the first basis vector is the lowest weight vector $|\mblw \rangle$, whereas the last basis vector is the  highest weight vector $|\mbhw \rangle$. Let us see some particular simple examples.

\subsection{U(2)-spin matrices for $M=1$ and $L=1$}

The general formulas \eqref{coef} for the Gelfand-Tsetlin basis vectors 
\begin{align}
	|1\ra=\left|
	\begin{matrix} 
		& 1 && 0 &  \\
		&& 0 & 
	\end{matrix}
	\right\rangle
	,\quad 
	|2\ra=\left|
	\begin{matrix} 
		& 1 && 0 &  \\
		&& 1 & 
	\end{matrix}
	\right\rangle\,,
\end{align}
give the $\rmu(2)$-spin operator matrices
\begin{equation}
	S_{11}=
	\begin{pmatrix}
		0&0\\
		0&1
	\end{pmatrix}
	,\quad
	S_{12}=
	\begin{pmatrix}
		0&0\\
		1&0
	\end{pmatrix}
	,\quad
	S_{21}=
	\begin{pmatrix}
		0&1\\
		0&0
	\end{pmatrix}
	,\quad
	S_{22}=
	\begin{pmatrix}
		1&0\\
		0&0
	\end{pmatrix}\:.
\end{equation}
These matrices are in one-to-one correspondence  to the $\text{SU(2)}$ Pauli matrices ($j=1/2$), $\sigma_z=S_{22}-S_{11}$, $\sigma_+=S_{21}$, $\sigma_-=S_{12}\,$ plus identity.

\subsection{U(2)-spin matrices for $M=1$ and $L=2$}

\begin{align}
	|1\ra=\left|
	\begin{matrix} 
		& 2 && 0 &  \\
		&& 0 & 
	\end{matrix}
	\right\rangle
	,\quad 
	|2\ra=\left|
	\begin{matrix} 
		& 2 && 0 &  \\
		&& 1 & 
	\end{matrix}
	\right\rangle
	,\quad 
	|3\ra=\left|
	\begin{matrix} 
		& 2 && 0 &  \\
		&& 2 & 
	\end{matrix}
	\right\rangle\,.
\end{align}

\begin{equation}
	S_{11}=
	\begin{pmatrix}
		0&0&0\\
		0&1&0\\
		0&0&2
	\end{pmatrix}
	,\quad
	S_{12}=
	\begin{pmatrix}
		0&0&0\\
		\sqrt{2}&0&0\\
		0&\sqrt{2}&0
	\end{pmatrix}
	,\quad
	S_{21}=
	\begin{pmatrix}
		0&\sqrt{2}&0\\
		0&0&\sqrt{2}\\
		0&0&0
	\end{pmatrix}
	,\quad
	S_{22}=
	\begin{pmatrix}
		2&0&0\\
		0&1&0\\
		0&0&0
	\end{pmatrix}\:.
\end{equation}
These matrices are related to the $\text{SU(2)}$ spin $j=1$ matrices, $J_z=\frac{1}{2}\left(S_{22}-S_{11}\right)$, $J_+=S_{21}$, $J_-=S_{12}\,$.

\subsection{U(4)-spin matrices for $M=2$ and $L=1$}

\begin{align}&
	|1\ra=\left|
	\begin{matrix}
		1 &&  1 && 0 && 0 & \\ 
		& 1 && 0 && 0 &&  \\
		&& 0 && 0 &&& \\
		&&& 0 &&&&
	\end{matrix}
	\right\rangle
	,\quad 
	|2\ra=\left|
	\begin{matrix}
		1 &&  1 && 0 && 0 & \\ 
		& 1 && 0 && 0 &&  \\
		&& 1 && 0 &&& \\
		&&& 0 &&&&
	\end{matrix}
	\right\rangle
	,\quad 
	|3\ra=\left|
	\begin{matrix}
		1 &&  1 && 0 && 0 & \\ 
		& 1 && 0 && 0 &&  \\
		&& 1 && 0 &&& \\
		&&& 1 &&&&
	\end{matrix}\right\rangle,
	\\
	&
	|4\ra=\left|
	\begin{matrix}
		1 &&  1 && 0 && 0 & \\ 
		& 1 && 1 && 0 &&  \\
		&& 1 && 0 &&& \\
		&&& 0 &&&&
	\end{matrix}
	\right\rangle
	,\quad 
	|5\ra=\left|
	\begin{matrix}
		1 &&  1 && 0 && 0 & \\ 
		& 1 && 1 && 0 &&  \\
		&& 1 && 0 &&& \\
		&&& 1 &&&&
	\end{matrix}
	\right\rangle
	,\quad 
	|6\ra=\left|
	\begin{matrix}
		1 &&  1 && 0 && 0 & \\ 
		& 1 && 1 && 0 &&  \\
		&& 1 && 1 &&& \\
		&&& 1 &&&&
	\end{matrix}
	\right\rangle.
	\nonumber
\end{align}

\begin{equation*}
	S_{11}=
	\begin{pmatrix}
		0&0&0&0&0&0\\
		0&0&0&0&0&0\\
		0&0&1&0&0&0\\
		0&0&0&0&0&0\\
		0&0&0&0&1&0\\
		0&0&0&0&0&1\\
	\end{pmatrix}
	,\quad
	S_{22}=
	\begin{pmatrix}
		0&0&0&0&0&0\\
		0&1&0&0&0&0\\
		0&0&0&0&0&0\\
		0&0&0&1&0&0\\
		0&0&0&0&0&0\\
		0&0&0&0&0&1\\
	\end{pmatrix}
	,\quad
	S_{33}=
	\begin{pmatrix}
		1&0&0&0&0&0\\
		0&0&0&0&0&0\\
		0&0&0&0&0&0\\
		0&0&0&1&0&0\\
		0&0&0&0&1&0\\
		0&0&0&0&0&0\\
	\end{pmatrix}
	,\quad
	S_{44}=
	\begin{pmatrix}
		1&0&0&0&0&0\\
		0&1&0&0&0&0\\
		0&0&1&0&0&0\\
		0&0&0&0&0&0\\
		0&0&0&0&0&0\\
		0&0&0&0&0&0\\
	\end{pmatrix},
\end{equation*}
\begin{equation}
	S_{12}=
	\begin{pmatrix}
		0&0&0&0&0&0\\
		0&0&0&0&0&0\\
		0&1&0&0&0&0\\
		0&0&0&0&0&0\\
		0&0&0&1&0&0\\
		0&0&0&0&0&0\\
	\end{pmatrix}
	,\quad
	S_{23}=
	\begin{pmatrix}
		0&0&0&0&0&0\\
		1&0&0&0&0&0\\
		0&0&0&0&0&0\\
		0&0&0&0&0&0\\
		0&0&0&0&0&0\\
		0&0&0&0&1&0\\
	\end{pmatrix}
	,\quad
	S_{34}=
	\begin{pmatrix}
		0&0&0&0&0&0\\
		0&0&0&0&0&0\\
		0&0&0&0&0&0\\
		0&1&0&0&0&0\\
		0&0&1&0&0&0\\
		0&0&0&0&0&0\\
	\end{pmatrix}.
\end{equation}
The rest of the $S_{ij}$ matrices can be obtained using the commutation relations in the equation \eqref{recurrence}.

\section{The case of non-rectangular Young tableaux}\label{appsec}

For a Young tableau of general shape $m_N=[m_{1N},\dots,m_{NN}]$, the HW state \eqref{highestweight} is generalized to
\be
|\mbhw\ra=\mathcal{N}(m_N) \Delta_{1}^{m_{1N}-m_{2N}} \Delta_{12}^{m_{2N}-m_{3N}}\dots   \Delta_{1,\dots,N-1}^{m_{N-1,N}-m_{NN}}\Delta_{1,\dots,N}^{m_{NN}}|0\ra_\mathrm{F},\label{highestweightN}
\ee
where $\Delta_{1,\dots,n}=\det(A_{\{1,\dots,n\}}^\dag)$  are leading (corner) principal minors of order $n$  of $A^\dag$ (for $M=N$, in general), like in  \eqref{minorsprod};  
$\mathcal{N}(m_N)$ denotes a normalizing factor. This HW state satisfies the 
HW conditions:
\bea
\Lambda_{\mu\nu}|\mbhw \rangle&=&
m_{\mu N}\delta_{\mu\nu} |\mbhw\rangle\,,\quad \mu\le \nu \label{p11}\,,\\ 
S_{ij}|\mbhw \rangle&=& m_{iN}\delta_{ij} |\mbhw \rangle\,,\quad i\leq j\,.
\label{p22}
\eea
If all components of $m_N$ are different, that is, $m_{1N}>m_{2N}>\dots>m_{NN}$, then all leading principal minors $\Delta_I$ of $A^\dag$ are present in the product \eqref{highestweightN} and 
the HW state $|\mbhw \rangle$ is only invariant under $\rmu(1)^N\subset \rmu(N)$ (all internal/flavor states $i=1,\dots,N$ are occupied). In this case, the ferromagnetic order parameter associated to the symmetry breaking is labeled 
by $\mathrm{dim}_\mathbb{C}[\rmu(N)/\rmu(1)^N]=N(N-1)/2$ complex parameters $z_{ij}\in\mathbb{C}, i>j=1,\dots,N-1$, parameterizing the coset (\emph{flag manifold}) $\mathbb{F}_{N-1}=\rmu(N)/\rmu(1)^N$. See e.g. \cite{CALIXTO1} 
for the Bruhat-Iwasawa decomposition in this case.

\bibliographystyle{apsrev4-1}

\bibliography{/home/usuario/MEGA_Genfimat/Bibliografia/bibliografia.bib}


\end{document}